\documentclass[3p,review]{elsarticle}

\usepackage{graphicx}
\usepackage{lmodern}

\usepackage{lineno,amstext,amssymb,bm}
\usepackage{mathtools} % For underbrace
\usepackage[dvipsnames]{xcolor} % To set text color
\definecolor{light-gray}{gray}{0.95}

\usepackage{amsthm}

\theoremstyle{remark}
\newtheorem*{remark}{Remark}

\usepackage{enumitem}
\usepackage{lipsum}

\usepackage{booktabs}
\usepackage{tabu}
\usepackage{listings}

\definecolor{codegreen}{rgb}{0,0.6,0}
\definecolor{codegray}{rgb}{0.5,0.5,0.5}
\definecolor{codepurple}{rgb}{0.58,0,0.82}
\definecolor{backcolour}{rgb}{0.95,0.95,0.92}

\lstdefinestyle{mystyle}{
  commentstyle=\color{codegreen},
  keywordstyle=\color{blue},
  numberstyle=\tiny\ttfamily\color{codegray},
  stringstyle=\color{codepurple},
  basicstyle=\ttfamily\tiny,
  breakatwhitespace=false,
  breaklines=true,
  captionpos=b,
  keepspaces=false,
  numbers=left,
  numbersep=6pt,
  showspaces=false,
  showstringspaces=false,
  showtabs=false,
  tabsize=2
}

\newcommand{\disp}{\boldsymbol{u}}
\newcommand{\dmg}{d}
\newcommand{\xvec}{\boldsymbol{x}}
\newcommand{\dd}[1]{\mathrm{d}#1}
\newcommand{\straint}{\boldsymbol{\epsilon}}
\newcommand{\stresst}{\boldsymbol{\sigma}}

\newcommand{\fb}{\boldsymbol{F_B}}
\newcommand{\ft}{\boldsymbol{F_T}}

\usepackage{float}
\usepackage{natbib}
\usepackage[bookmarks=true,colorlinks=true,linkcolor=blue,citecolor=red,backref=page]{hyperref}
\makeatletter
\AtBeginDocument{\def\@citecolor{red}}
\makeatother

\makeatletter
\def\cl@chapter{\@elt {theorem}}
\makeatother %https://tex.stackexchange.com/questions/108394/tex-capacity-exceeded-sorry-input-stacksize-5000
\usepackage[nameinlink,capitalize]{cleveref}
\usepackage{algorithm, algpseudocode}

\makeatletter
\newcounter{algorithmicH}% New algorithmic-like hyperref counter
\let\oldalgorithmic\algorithmic
\renewcommand{\algorithmic}{%
  \stepcounter{algorithmicH}% Step counter
  \oldalgorithmic}% Do what was always done with algorithmic environment
\renewcommand{\theHALG@line}{ALG@line.\thealgorithmicH.\arabic{ALG@line}}
\makeatother

\begin{document}
    \begin{frontmatter}

    \title{A 55-line code for large-scale parallel topology optimization in 2D and 3D}

    %% Group authors per affiliation:
    \author[my_main_address]{Abhinav Gupta}
    \ead{agupta1[at]ce.iitr.ac.in}

    \author[my_main_address]{\texorpdfstring{Rajib Chowdhury\corref{corresponding_author}\fnref{footnote_1}}{}}
    \cortext[corresponding_author]{Corresponding author}
    \ead{rajibfce[at]iitr.ac.in, rajibchowdhury[at]ce.iitr.ac.in}
    \fntext[footnote_1]{Tel.: (+91) 1332-285612}

    \author[my_main_address]{\texorpdfstring{Anupam Chakrabarti\fnref{footnote_2}}{}}
    \fntext[footnote_2]{Tel.: (+91) 1332-285844}
    \ead{anupam.chakrabart[at]ce.iitr.ac.in}

    \author[timon_address]{Timon Rabczuk}
    \ead{timon.rabczuk[at]uni-weimar.de}

    \address[my_main_address]{Department of Civil Engineering,
    Indian Institute of Technology Roorkee, India}

    \address[timon_address]{Institute of Structural Mechanics, Bauhaus-Universität Weimar, 99423 Weimar, Germany}

\begin{abstract}

    This paper presents a 55-line code written in python for 2D and 3D topology optimization (TO) based on the open-source finite element computing software (FEniCS), equipped with various finite element tools and solvers. PETSc is used as the linear algebra back-end, which results in significantly less computational time than standard python libraries. The code is designed based on the popular solid isotropic material with penalization (SIMP) methodology. Extensions to multiple load cases, different boundary conditions, and incorporation of passive elements are also presented. Thus, this implementation is the most compact implementation of SIMP based topology optimization for 3D as well as 2D problems.

    Utilizing the concept of Euclidean distance matrix to vectorize the computation of the weight matrix for the filter, we have achieved a substantial reduction in the computational time and have also made it possible for the code to work with complex ground structure configurations. We have also presented the code's extension to large-scale topology optimization problems with support for parallel computations on complex structural configuration, which could help students and researchers explore novel insights into the TO problem with dense meshes. \ref{sec:appenix_a} contains the complete code, and the website: \url{https://github.com/iitrabhi/topo-fenics} also contains the complete code.
\end{abstract}


\end{frontmatter}

    \section{Introduction}
%\textcolor{blue}{What is the definition of topology optimization? Who initiated the study in this field?}

The primary purpose of structural topology optimization is to generate an optimal design - by removing material from the design domain - which should effectively comply with its intended design objectives and, at the same time satisfies the constraints imposed upon it. The development of high-performance structural topology optimization methods would result in low-cost, lightweight, and high-performance structures. For the first time, \citet{bendsoeGeneratingOptimalTopologies1988} presented the theory for topology optimization of continuum structures. The initial works were based on the homogenization or micro-structure approach. This approach provides criteria for topology optimization using composite materials to describe different material properties where each element is a micro-structure for a given homogeneous, isotropic material. However, the determination and evaluation of optimal micro-structures and their orientations are cumbersome \citep{bendsoeGeneratingOptimalTopologies1988}.

In contrast, the Solid Isotropic Material with Penalization(SIMP) approach proposed by \citet{bendsoeOptimalShapeDesign1989,zhouCOCAlgorithmPart1991,mlejnekAspectsGenesisStructures1992} suggest a relation between the design density variable and the material property based on the power-law, which results in improving the convergence of the solution.  Because of the ease of implementation and straightforward description of the design density variable, this approach has grown in popularity among the research community. It is generally the preferred way to learn and understand the working philosophy of topology optimization. Gradient-based optimization algorithms have become the preferred way to solve the smooth and differentiable topology optimization problem represented by the SIMP approach\citep{bendsoeTopologyOptimizationTheory2013, sigmund99LineTopology2001, sigmundDesignCompliantMechanisms1997, sigmundMorphologybasedBlackWhite2007,sigmundNumericalInstabilitiesTopology1998,sigmundTopologyOptimizationApproaches2013b}. Topology optimization method has also been recently studied with the iso-geometric approach \citep{kangIsogeometricTopologyOptimization2016,nguyenThreedimensionalTopologyOptimization2020,ghasemiLevelsetBasedIGA2017,hamdiaNovelDeepLearning2019a} and mixed FEM approach \citep{nanthakumarTopologyOptimizationFlexoelectric2017, nanthakumarTopologyOptimizationPiezoelectric2016}

%\textcolor{blue}{What are the significant contributions to the code base for topology optimization}

The classical 99-line MATLAB code for topology optimization of continuum structures \citep{sigmund99LineTopology2001} lead to a plethora of efficient implementations of topology optimization in various programming languages utilizing different methods. The initial code - which was based on SIMP method - has helped many researchers to develop their own implementations with SIMP \citep{kharmandaReliabilitybasedTopologyOptimization2004,suresh199lineMatlabCode2010,andreassenEfficientTopologyOptimization2011,schmidt2589LineTopology2011,talischiPolytopMATLABImplementation2012,liuEfficient3DTopology2014}, Level Set\citep{wangLevelSetMethod2003,challisDiscreteLevelsetTopology2010,allairegAllaire20092d2009,otomoriMATLABCodeLevel2014,wei88lineMATLABCode2018, herreroImplementationLevelSet2013,ortigosaNewStabilisationApproach2019,perezLevelSetMethod2012,ghasemiLevelsetBasedIGA2017,ghasemiMultimaterialLevelSetbased2018,luTopologyOptimizationAcoustic2013}, reaction-diffusion{seongReactiondiffusionEquationBased2018a,jeongStructuralDesignConsidering2019}, Bi-Directional Evolutionary Structural Optimization\citep{zuoSimpleCompactPython2015,zhouDesignFabricationBiphasic2012,martinez-frutosEvolutionaryTopologyOptimization2018} and  Moving Morphable Component\citep{zhangNewTopologyOptimization2016}. Further \citet{schmidt2589LineTopology2011,aageGigavoxelComputationalMorphogenesis2017a, aageParallelFrameworkTopology2013, aageTopologyOptimizationUsing2015, amirMultigridCGEfficientTopology2014} have proposed efficient and parallelizable codes to carry out large scale topology optimization simulations. Recently a new generation 99 line MATLAB code was presented by \citet{ferrariNewGeneration992020}. Readers are referred to \citep{wei88lineMATLABCode2018} for a comprehensive review of all the educational codes available in literature.  These codes have proved to be an integral part of the learning process for many researchers who wish to understand and implement topology optimization in their work.

%\textcolor{blue}{What are the challenges involved in modeling topology optimization with existing codal implementations.}

The attractive primary property of these codes is that they are easy to understand and modify. However, to achieve readability and compactness, most of the proposed codes are limited to specific problems. Works by \citet{zuoSimpleCompactPython2015, chen213lineTopologyOptimization2019} have helped expand the applicability of the codes to complex problems using ABAQUS and ANSYS. These works have made it possible to take advantage of the linear/nonlinear, static/dynamic FEA capacities, and meshing techniques present in the commercial package but then loose on the flexibility provided by custom implementations. Further, this limits the extension of these codes by the researches who do not have access to the commercial packages.

%\textcolor{blue}{What do you propose and what are the benefits of your proposal.}

In this paper, we have presented a 55-line code for carrying out SIMP based topology optimization in FEniCS. The intention of this code is to stay true to the the original so that researchers and students who are already familiar with the topic find it easy to understand an implement in their workflow. We propose the use of Euclidean distance matrices to vectorize the computation of distance matrices for filter. This results in  substantial reduction in computational time for evaluation of the filter. Moreover, this method supports complex structural configurations. We also present the extension of the code to large scale complex engineering structure with support for parallelization. Thus, this compact code could help students and researchers learn and understand the concept of TO on their laptops and then extend the same to handle complex 3D structures on large high-performance computers.

%\textcolor{blue}{Organization:}

This manuscript is structured as follows. \cref{sec:formulation} addresses the theoretical aspects of the variational formulation of the linear elasticity problem and discusses the SIMP approach. \cref{sec:nuerical_approximation} presents the mathematical model of the topology optimization problem. We discuss the continuous and discontinuous Lagrange elements used for discretization and then present the problem's weak form. We then discuss the optimization approach, the mesh independency filter, and the Euclidean distance matrix (used to vectorize the loops for calculating the distance matrix for the mesh independency filter). In \cref{sec:fenics}, we present a detailed step by step description of our FEniCS implementation. \cref{sec:extensions} includes the model extensions of the code to a 3D problem, different load cases, different boundary conditions, and inclusion of passive elements. Finally,  we summarize the findings, in \cref{sec:conclusion}.

    \section{Topology optimization formulation}
\label{sec:formulation}

\begin{figure}
    \centering
    \includegraphics{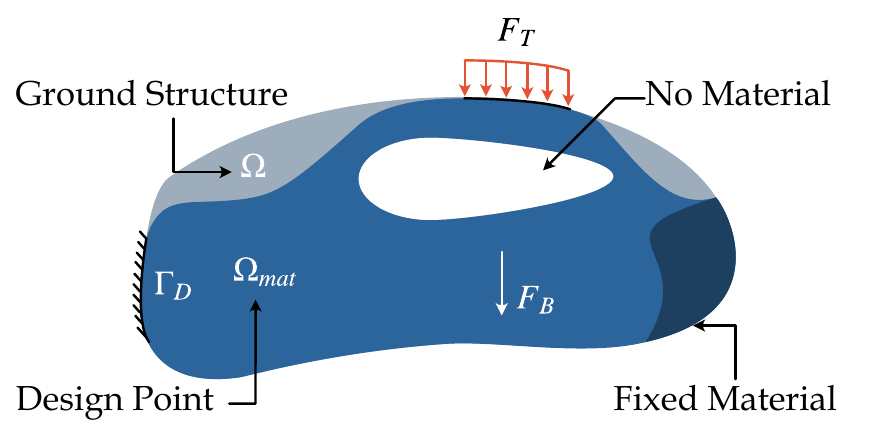}
    \caption{Structural topology optimization problem definition. The ground structure is chosen to allow for a definition of applied loads and boundary conditions. It could consist of zones with no material or zones with fixed material.}
    \label{fig:domain}
\end{figure}

We consider an arbitrarily shaped linear elastic body $\Omega_{mat}$ which is part of a large reference domain $\Omega \subset \mathbb{R}^{\delta}, {\delta} = 2,3$, having boundary $\Gamma$, as illustrated in \cref{fig:domain}. The reference domain $\Omega$ (ground structure) is chosen to allow for a definition of applied loads and boundary conditions. The displacement at a point $\xvec \in \Omega$ is represented by $\disp(\xvec)$.  The Dirichlet and Neumann boundaries of the domain $\Omega$ are represented by $\Gamma_D$ and $\Gamma_N$, respectively. Let $\fb \subset \mathbb{R}^{\delta}, {\delta} = 2,3$ represent the body force vector and $\ft \subset \mathbb{R}^{\delta}, {\delta} = 2,3$ represent the traction force on the on the traction part $\Gamma_T \subset \Gamma$, of the boundary. The ground structure could consist of zones with no material(such as holes), or zones with fixed material. FEniCS works based on the variational formulation of the problem, and thus in this section we discuss the variational approach to topology optimization problem.

\subsection{Governing equations and weak form
}
Elastic strain energy is the energy stored at the time of deformation inside an elastic body. When an isotropic linear elastic solid is deformed under small strain condition, the strain energy density function $\psi_0(\straint)$ is defined as:
\begin{equation}
        \psi_0(\straint)  = \frac{\lambda}{2} (tr[\straint])^2 + \mu tr[\straint^2]
        \label{eq:str_energy}
\end{equation}
where $\mu, \lambda > 0$ are the Lam\'{e} parameters, $\epsilon = \frac{1}{2}(\nabla{\disp} + \nabla{\disp}^T) = sym(\nabla \disp)$, is defined as the symmetric part of the tensor $\nabla{\disp}$. The total strain energy is thus defined as:
\begin{equation}\label{eq:elastic_energy}
    SE = \int_\Omega{\psi_0(\bm{\epsilon}) \dd{\Omega}}
\end{equation}

%\subsection{Work Potential}
The work potential is defined as the potential of the external loads, which is the negative of the work done on the system. The work potential is given by
\begin{equation}
    WP = -\int_{\Omega}{\disp . \fb}\dd{\Omega} -\int_{\Gamma}{\disp . \ft}\dd{\Gamma}
\end{equation}

%\subsection{Potential Energy}

The potential energy of the system is defined as the sum of elastic energy and work potential. The total potential energy for a deformed body is thus given by:
\begin{equation}\label{eq:potential_energy}
U = \int_\Omega{\psi_0(\bm{\epsilon}) \dd{\Omega}}-\int_{\Omega}{\disp . \fb}\dd{\Omega} -\int_{\Gamma}{\disp . \ft}\dd{\Gamma}
\end{equation}

%\subsection{Variational form}
For calculating displacement field the principle of minimum potential energy is used. Thus, by taking the first variation of $U$ and then applying the fundamental lemma of the calculus of variation, we get:

\textbf{Governing differential equation:}

\begin{equation}\label{eq:momentum_eq}
    \nabla \cdot \stresst\ = -\fb \quad in \quad \Omega
\end{equation}

\textbf{Natural boundary condition:}\\
\begin{equation}
\begin{split}
    (\stresst . \bm{n}) =\ft \quad on \quad \Gamma
\end{split}
\end{equation}

%\subsection{SIMP Method}
In the SIMP method, we introduce a  pseudo density $d \in [0, 1]$ into  \cref{eq:momentum_eq} to model the distribution of the material within the ground structure.
\begin{equation}
\label{eq:simp_fn}
 \dmg^p\ \nabla . \stresst = \fb
\end{equation}
In \cref{eq:simp_fn}, $p$ is the penalty parameter that helps in enforcing a $0/1$ density distribution by significantly reducing the effect of intermediate values. Based on the power-law approach we could now define the topology optimization problem, where the objective is to minimize the strain energy and the volume is constrained.
\begin{equation}
\label{eq:simp}
\begin{aligned}
& \underset{\dmg}{\text{minimize}}
& & f(\dmg) = \dmg^p U \\
& \text{subject to}
& & g(\dmg) = \int_{\Omega} \dmg . V_0 \dd{\Omega} - k. V_0 \leq 0\\
& & & \forall \dmg \in \mathbb{D}, \mathbb{D} = \{\dmg \in L^2(\Omega):\dmg_{min}<\dmg\leq 1\}
\end{aligned}
\end{equation}
In \cref{eq:simp}, $U$ is the strain energy of the domain and $V_0$ is the initial volume. $k$ is the prescribed volume fraction.

    \section{Numerical Approximation}
\label{sec:nuerical_approximation}
The linear elasticity model, as described by the differential equation (refer \cref{eq:momentum_eq}), is discretized by the finite element method. We start with a discussion on weak form and the discrete form of the system. Then we discuss the optimization algorithm and the filter used in the method to avoid the checkerboard pattern.
    \subsection{Spatial discretization}
\label{sec:discretization}
\begin{figure}
    \centering
    \includegraphics{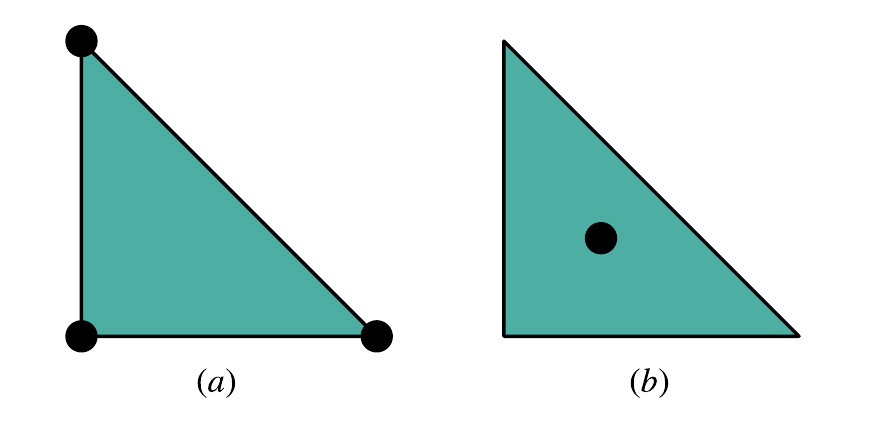}
    \caption{Finite elements - 2D. (a)three noded triangular Lagrangian element with two degrees of freedom per node used to model the displacement vector. (b) one noded discontinuous Lagrangian element used to model the density parameter.}
    \label{fig:elements}
\end{figure}

\begin{figure}
    \centering
    \includegraphics{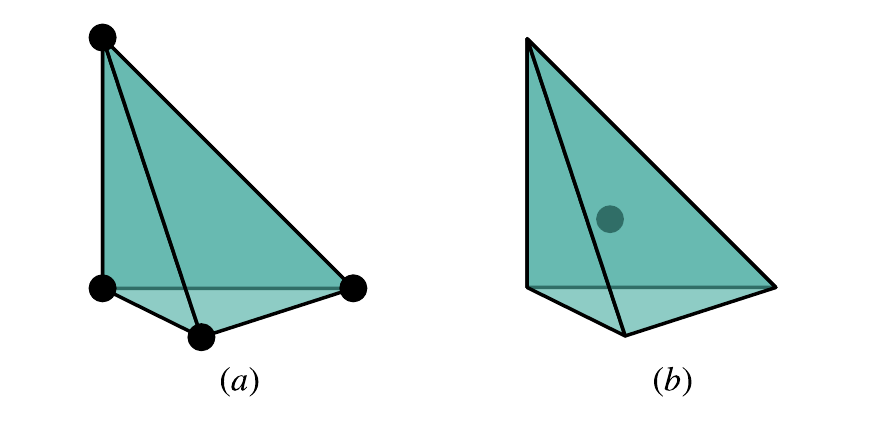}
    \caption{Finite elements - 3D. (a) four noded tetrahedral Lagrangian element with three degrees of freedom per node used to model the displacement vector. (b) one noded discontinuous Lagrangian element used to model the density parameter.}
    \label{fig:elements-3d}
\end{figure}

We use first degree $C^0$ Lagrangian finite elements to discretize the domain and the primary field variable. The density parameter is discretized with the help of zeroth degree, $C^{-1}$ discontinuous Galerkin elements. In the 2 dimensional cases, the three noded triangular Lagrangian finite elements have two degrees of freedom per node, whereas the discontinuous Lagrangian element of degree zero have a single value per element and are represented as shown in \cref{fig:elements}. For the 3 dimensional case, the four noded tetrahedral Lagrangian finite element have three degrees of freedom per node, whereas the discontinuous Lagrangian elements of degree zero have a single value per element and are represented as shown in \cref{fig:elements-3d}.

\begin{remark}
Triangular and tetrahedral elements are used for the current implementation as they are well supported in FEniCS, whereas the support for quadrilateral and hexahedral elements, even though present is somewhat limited. For simple formulations, user could switch to  quadrilateral and hexahedral elements to reduce the computational expense.
\end{remark}

\subsubsection{Weak form}
The weak form of the system is derived by first multiplying the governing differential equation (refer  \cref{eq:momentum_eq}) by admissible test function ($\bm{v}$) and then modification of the equations with the application of Green's theorem. The weak form of the system is defined as: Find $\disp \in \bm{U}$ such that
\begin{equation}
\label{eq:weak}
    \begin{split}
        \int_\Omega \dmg^p \stresst(\disp):\nabla \bm{v} \ d\Omega = 0 \quad \forall \bm{v} \in \widehat{\bm{U}}
    \end{split}
\end{equation}

In \cref{eq:weak}, one usually uses only the symmetric part of the test function (virtual strain, $\bm{\epsilon}(\bm{v}) = 1/2 (\nabla \bm{v}+\nabla \bm{v}^T$) as $\int_\Omega \stresst(\disp): (\nabla \bm{v})_{skewsym} \ d\Omega = 0$.
\begin{equation}
\label{eq:weak-sym}
    \begin{split}
        \int_\Omega \dmg^p \stresst(\disp):\bm{\epsilon}(\bm{v}) \ d\Omega = 0 \quad \forall \bm{v} \in \widehat{\bm{U}}
    \end{split}
\end{equation}

The trial ($\bm{U}$) and test ($\widehat{\bm{U}}$) spaces are defined as
\begin{equation}
\label{eq:space}
    \begin{split}
        (\bm{U}, \widehat{\bm{U}}) = \{(\disp, \bm{v}) \in H^1(\Omega): \disp=\overline{\disp} \ \&\  \bm{v}=0 \ on \ \Gamma_D\}
    \end{split}
\end{equation}

In \cref{eq:space}, $H^1(\Omega)$ denotes the usual Hilbert space over the domain $\Omega$ and $\overline{\disp}$ is the prescribed displacement.

\subsubsection{Discrete form}
The weak form of the system as presented by \cref{eq:weak} is discretized by using $C^0$ Lagrangian elements for displacement vector and $C^{-1}$ discontinuous Lagrangian elements for density. The discrete form of the problem is thus defined as: Find $(\bm{u_h}) \in \bm{U_h}\subset (\bm{U})$ such that
\begin{equation}\label{eq:fe_form}
    \begin{split}
        \int_\Omega \dmg_h^p \stresst(\bm{u_h}):\bm{\epsilon}(\bm{v}) \ d\Omega = 0 \quad \forall \bm{v} \in \widehat{\bm{U}}_h\subset \widehat{\bm{U}}
    \end{split}
\end{equation}

We follow the Bubnov-Galerkin formulation, where the test and trial functions belong to the same function space. Continuous field variables are thus approximated as:
\begin{equation}
\begin{split}
    \disp(\xvec)\approx \disp_h(\xvec) = \sum_{i=1}^n \bm{N}_i \disp_i
\end{split}
\end{equation}

where $n$ denotes the number of nodes in the mesh, $\disp_i$ and $N_i$ are the displacement vector and basis function corresponding to $i^{th}$ node, respectively.

    \subsection{Optimization algorithm}
    The optimization problem as described in \cref{eq:simp} is solved using the standard optimality criteria (OC) method. The OC method states that convergence is achieved when the KKT condition is satisfied \citep{liuEfficient3DTopology2014},

    \begin{equation}
    \label{eq:kkt}
    \frac{\partial f(\dmg)}{\partial \dmg_{e}}+\lambda \frac{\partial g(\dmg)}{\partial \dmg_{e}}=0
    \end{equation}

In \cref{eq:kkt} $\lambda$ is the Lagrange multiplier associated with the constraint $g(\dmg)$. This optimality condition can be expressed as $B_e = 1$, where
\begin{equation}
B_{e}=-\frac{\partial f(\dmg)}{\partial \dmg_{e}}\left(\lambda \frac{\partial g(\dmg)}{\partial \dmg_{e}}\right)^{-1}
\end{equation}

According to \citep{bendsoeOptimizationStructuralTopology1995} a heuristic updating scheme for the design variables can be formulated as:
\begin{equation}
\label{eq:update}
d_{e}^{\mathrm{new}}=\left\{\begin{array}{ll}
\max \left(d_{min}, d_{e}-m\right) & \text { if } d_{e} B_{e}^{\eta} \leq \max \left(d_{min}, d_{e}-m\right) \\
\min \left(1, d_{e}+m\right) & \text { if } d_{e} B_{e}^{\eta} \geq \min \left(1, d_{e}+m\right) \\
d_{e} B_{e}^{\eta} & \text { otherwise }
\end{array}\right.
\end{equation}

In \cref{eq:update}, $m$ is a positive move-limit, and $\eta$ is a numerical damping coefficient. The choice of $m = 0.2$ and $\eta = 0.5$ is recommended \citep{bendsoeOptimizationStructuralTopology1995,sigmund99LineTopology2001}.
\subsubsection{The bisection method}
We use the bisection method to find the Lagrange multiplier $\lambda$. The Lagrange multiplier should satisfy the constraint function $g(\dmg)$, i.e. the new value of density parameter evaluated with \cref{eq:update} should satisfy the condition
\begin{equation}
g(d(\lambda)) = 0
\end{equation}
Numerically, this could be achieved by the bisection method in which we assume a lower bound and upper bound on the value of $\lambda$ and calculate the value of constraint function at mid-point of that interval. If the convergence is satisfactory (i.e. the length of the interval is small), then stop the iteration and return mid-point as solution. Otherwise, examine the sign of the constraint function and update the lower-bound with mid-point if the sign is positive else update the upper-bound with mid-point value.

The sensitivity of the objective function is defined as
\begin{equation}
\frac{\partial f(d)}{\partial d} = -p d^{p-1}U
\end{equation}
The derivative of the volume constraint $g(d)$ in \cref{eq:simp} with respect to the design variable $d$ is given
\begin{equation}
\frac{\partial g(d)}{\partial d}=   \int_{\Omega} V_0 \dd{\Omega}
\end{equation}

\subsubsection{Mesh-independency filter}
\label{sec:mif}
    The solution of topology optimization problems is mesh-dependent. To circumvent this issue, researchers have proposed sensitivity filters that ensure mesh-independency.  The filter works by modifying the sensitivity of a specific element's objective function by taking the weighted average of the element sensitivities in a fixed neighborhood.
    \begin{equation}
    \frac{\partial (d_a E_a)}{\partial d_{a}}=\frac{1}{\sum_{b=1}^{N} {W}_{b}} \sum_{b=1}^{N} {W}_{b} \frac{\partial (d_b E_b)}{\partial d_{b}}
    \end{equation}
    \begin{equation}
    \label{eq:filter}
    \frac{\partial (E_a)}{\partial d_{a}}=\frac{1}{d_a \sum_{b=1}^{N} {W}_{b}} \sum_{b=1}^{N} {W}_{b} d_b\frac{\partial ( E_b)}{\partial d_{b}}
    \end{equation}

    \begin{figure}
    \centering
    \includegraphics{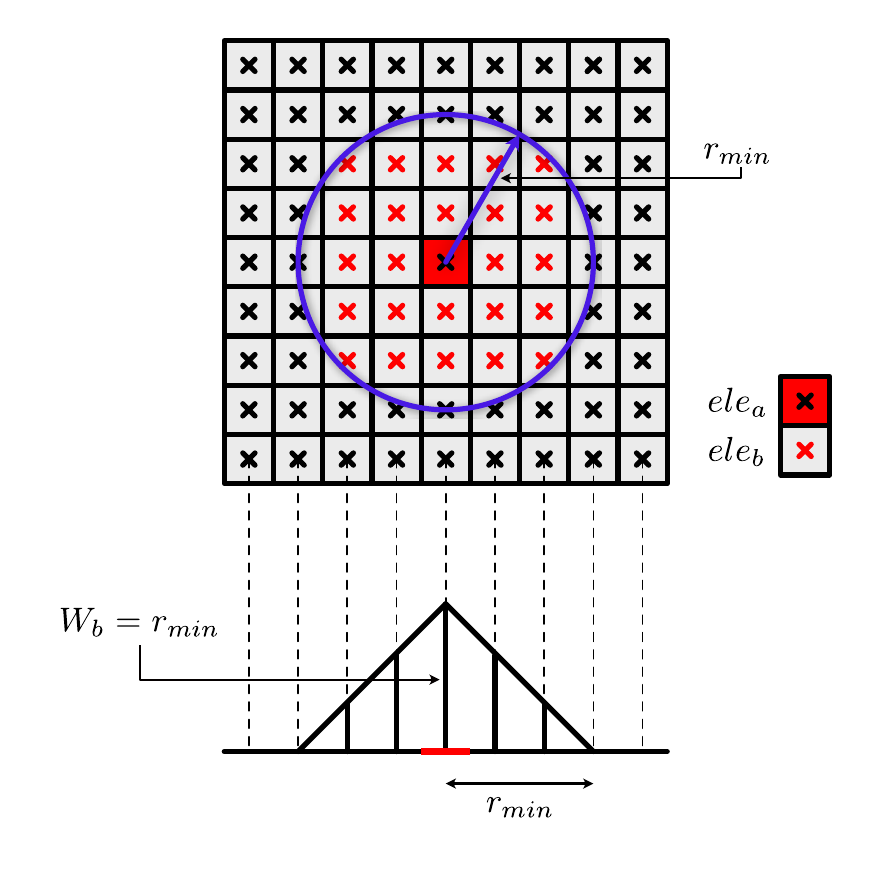}
    \caption{Mesh-independency filter. (top)Elements ($ele_b$) in the neighborhood of $ele_a$ considered for the density filtering. (bottom) Variation of the weight factor ($W_b$) with increasing distance from $ele_a$. Note that the weight factor is zero outside the filter zone marked by $r_{min}$}
    \label{fig:filter}
    \end{figure}

    Here the letter $'a'$ represents the element under consideration (marked as red in \cref{fig:filter}) and the letter $'b'$ represents the elements in the neighborhood of element $'a'$ (marked with red X marks in the \cref{fig:filter}). The elements in the neighborhood are identified as the ones whose center lies inside a circle of radius $r_{min}$ as shown in the \cref{fig:filter}. $W_b$ is the weight factor (also known as the convolution operator) assigned to the element $'b'$, which decays linearly with increasing distance from element $'a'$ and is defined as:

    \begin{equation}
    \label{eq:filter_cond}
    W_b = r_{min} - dist(b, a),\ \{b \in \mathbb{N} | dist(b,a) \leq r_{min})\} ,\ b=1,...,N.
    \end{equation}

    In \cref{eq:filter_cond}, $dist(b, a)$ is the distance operator that calculates the distance between the midpoints of the element $'b'$ and element $'a'$. The weight factor is zero outside the zone defined by the circle with a radius $r_{min}$.

    \subsection{Numerical implementation}
The numerical simulations are performed using the python interface of the open-source scientific computing platform FEniCS \citep{LoggMardalEtAl2012a,FEniCSAlnaesBlechta2015a, DOLFINLoggWells2010a,FFCKirbyLogg2006a,FFCOlgaardWells2010b,UFLAlnaesEtAl2012,FIATKirby2004a,UFCAlnaesLoggEtAl2009a,PetscBalay2010,bleyer2018numericaltours}. To solve the algebraic equations, FEniCS is configured with PETSc \citep{petsc-efficient,petsc-web-page,abhyankar2018petsc} as the linear algebra backend. The output of the topology optimization problem is written to a XDMF file and Paraview \citep{paraview} is used to visualize the output.

All the two dimensional analyses were performed on a Windows laptop equipped with an Intel Core i7 processor with a clock frequency of 2.2GHz having six cores and 16GB of RAM. The three dimensional analysis were performed on a workstation equipped with two
Intel Xeon Platinum 8260 M processor with a clock frequency of 2.4 GHz having 24 cores per processor amounting to a total of 48 cores and memory is 128 GB.

    \section{FEniCS implementation}
\label{sec:fenics}
    In this section, the 55-line FEniCS code for the energy minimization problem is explained in detail. Several assumptions were made to simplify the code in the previous compact codes (99-line and 88-line). Thus, the codes were limited to a certain kind of problem, and changing them to 3D would require significant modification of the codebase.

    \begin{figure}
    \centering
    \includegraphics[width=16cm]{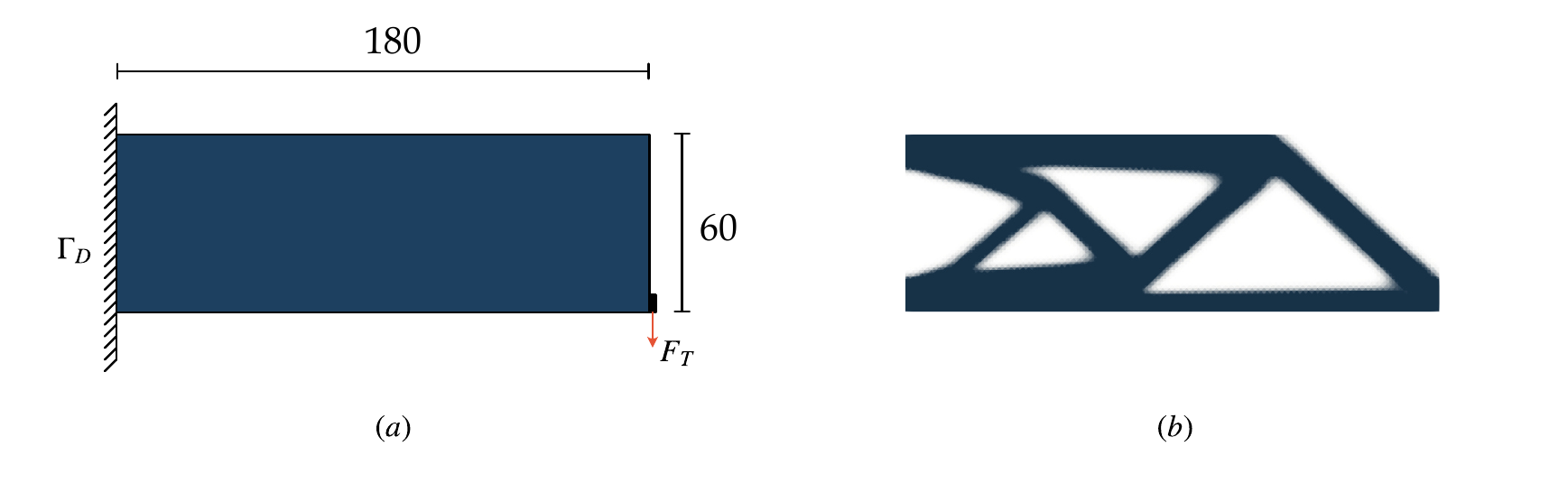}
    \caption{Cantilever beam problem. (a)The design domain, boundary conditions, and external load for the optimization of a cantilever beam. (b) Topologically optimized structure.}
    \label{fig:cantilever}
    \end{figure}

    The intention of the current implementation is to take advantage of the highly popular programming language python, which is known for its compactness and readability, and to create the most compact implementation of topology optimization by utilizing the open-source FEM package FEniCS. FEniCS allows us to describe the FEM problem in terms of variational principles, keeping the python implementation as close to the mathematical description as possible. This also helps us to improve the readability of the code when comparing it to the mathematical description. The code is called from the python prompt employing the following lines:

    \begin{lstlisting}[language=Python,numbers=none,xleftmargin=0.0cm,backgroundcolor=\color{light-gray},frame=tlbr,framesep=4pt]
from topopt import main
main(nelx=180, nely=60, volfrac=0.5, penal=3.0, rmin=2.0)
    \end{lstlisting}

    where \texttt{nelx} and \texttt{nely} are the number of elements in the horizontal and vertical direction, respectively, \texttt{volfrac} is the prescribed volume fraction $k$, \texttt{penal} is the penalization power $p$, \texttt{rmin} is the filter radius $r_{min}$. \cref{fig:domain}(a) shows the domain and boundary conditions for a cantilever beam as an example that would be used to describe the code.
\subsection{Header}
The first line imports all the key classes and methods necessary for solving a finite element variational form in FEniCS. The dolfin library is a part of the FENiCS package, which in itself is a wrapper for the C++ library DOLFIN that provides efficient and highly scalable C++ classes for finite element computing. Thus, the package benefits from the speed and efficiency of C++ while keeping the programming for the end-user in an easy to understand and highly readable programming language (python). The second line imports the two packages necessary for performing matrix operations.
    \begin{lstlisting}[language=Python,numbers=none,xleftmargin=0.0cm,backgroundcolor=\color{light-gray},frame=tlbr,framesep=4pt]
from dolfin import *
import numpy as np, sklearn.metrics.pairwise as sp
    \end{lstlisting}

\subsection{Function declaration}
In FEniCS we solve a problem by defining the variational formulation with Unified Form Language (UFL). This allows us to define the formulas in python that are very similar to the mathematical description and thus makes the code highly readable. The Cauchy stress tensor and the strain energy of the system are described as
    \begin{equation}
    \label{eq:stress}
        \sigma = \lambda\,\mathrm{tr}\,(\straint) I + 2\mu \straint
    \end{equation}
    \begin{equation}
        \label{eq:energy}
        \psi_0(\straint)  = \frac{\lambda}{2} (tr[\straint])^2 + \mu tr[\straint^2]
    \end{equation}
    In \cref{eq:stress,eq:energy}, strain ($\straint = \frac{1}{2}(\nabla{\disp} + \nabla{\disp}^T) = sym(\nabla \disp)$), is defined as the symmetric part of the  part of $\nabla \disp$. Note the very close correspondence between the mathematical formulas and their equivalent UFL descriptions. This is the core strength of FEniCS that we wish to exploit to the fullest to make the most compact code of topology optimization.
    \begin{lstlisting}[language=Python,numbers=none,xleftmargin=0.0cm,backgroundcolor=\color{light-gray},frame=tlbr,framesep=4pt]
sigma = lambda _u: 2.0 * mu * sym(grad(_u)) + lmbda * tr(sym(grad(_u))) * Identity(len(_u))
psi = lambda _u: lmbda / 2 * (tr(sym(grad(_u))) ** 2) + mu * tr(sym(grad(_u)) * sym(grad(_u)))
    \end{lstlisting}

\subsection{Finite element analysis}
Most of the codes available in the literature follow the approach of the matrix method, wherein they derive the stiffness matrix and force vectors and then code them into the programming language. FEniCS is based on the variational approach, where the user needs to specify the variational form with UFL and then use the built-in operators and methods to generate all the required matrices and vectors automatically. This process is initiated by creating function spaces over the mesh and then defining functions in that function space. The supports and loads are then defined based on the function space and functions. Finally, a linear variational problem and a solver are defined.

\subsubsection{Discretization the problem}\label{subsub:meshing}
We first start discretizing the problem by defining the domain under consideration with the built in function \texttt{RectangularMesh}. The mesh is made up of triangular elements. The parameters \texttt{nelx} and \texttt{nely} tell the program to divide the domain into \texttt{nelx} and \texttt{nely} unit rectangles in x and y directions respectively. Each of the unit rectangle is then divided into two triangles and the last argument decides the direction for the division of the rectangles.
\begin{lstlisting}[language=Python,numbers=none,xleftmargin=0.0cm,backgroundcolor=\color{light-gray},frame=tlbr,framesep=4pt]
mesh = RectangleMesh(Point(0, 0), Point(nelx, nely), nelx, nely, "right/left")
U = VectorFunctionSpace(mesh, "CG", 1)
D = FunctionSpace(mesh, "DG", 0)
u, v = TrialFunction(U), TestFunction(U)
u_new, density, density_old = Function(U), Function(D), Function(D)
\end{lstlisting}
As discussed in \cref{sec:discretization} the topology optimization problem consists of two primary variables, the displacement vector ($\disp$), and the density parameter ($d$) which is constant over the element and is discontinuous over the edges. Thus, to discretize the problem with finite element method we need to define two function spaces over the mesh. The first one is defined using the CG elements over the mesh and creating a vector function space of degree one. This implies that the element is a 3-noded linear triangle with two degrees of freedom at each node. For the density parameter we define a function space with DG elements. Once we have defined the function space we can define the test and trial functions over the space which are required to define the weak form of the variational problem. Then we define functions which are used to store the solution to the problem.
\subsubsection{Definition of supports}
The supports are first identified inside the mesh with the \texttt{CompiledSubdomain} method that takes in a condition in C++ syntax and then at compile time converts the expression into an efficient compiled C++ function. In the case of a cantilever beam fixed at x=0 the support is defined by the following command.
\begin{lstlisting}[language=Python,numbers=none,xleftmargin=0.0cm,backgroundcolor=\color{light-gray},frame=tlbr,framesep=4pt]
support = CompiledSubDomain("near(x[0], 0.0, tol) && on_boundary", tol=1e-14)
bcs = [DirichletBC(U, Constant((0.0, 0.0)), support)]
\end{lstlisting}
The C++ expression takes a point vector \texttt{x} as input, where the vector elements represent the x, y, and z coordinates. The support nodes are identified by a conditional statement based on the point vector. The Dirichlet boundary condition is then defined - over the nodes identified by the function \texttt{support} - by assigning a value of zero to the components of the displacement vector over the vector function space U $(\disp = 0\ on\ \Gamma_D)$.

\subsubsection{Definition of loads}
The load boundary is identified with the help of \texttt{load\char`_marker}. This is achieved in FEniCS with the help of \texttt{MeshFunction} method. We mark the mesh entities with integer markers. We then use those integers in the measures (\texttt{dx} and \texttt{ds}) to identify the region in our variational form. In UFL, \texttt{dx} represents the integral over the whole cell whereas \texttt{ds} represents the integral over the exterior facet of the cell. In our case suppose the load is applied over the bottom two elements of the mesh. This is achieved by first identifying the bottom two elements with the load marker and then updating the measure \texttt{ds} to account for the markers. After that we define the right hand $\bm{f}.\bm{v}\ ds$ with the command.
\begin{lstlisting}[language=Python,numbers=none,xleftmargin=0.0cm,backgroundcolor=\color{light-gray},frame=tlbr,framesep=4pt]
load_marker = MeshFunction("size_t", mesh, 1)
CompiledSubDomain("x[0]==l && x[1]<=2", l=nelx).mark(load_marker, 1)
ds = Measure("ds")(subdomain_data=load_marker)
F = dot(v, Constant((0.0, -1.0))) * ds(1)
\end{lstlisting}
\subsubsection{Defining the variational problem and solver}
The bilinear form of the systems as described by \cref {eq:fe_form} is defined first and then we define the \texttt{LinearVariationalProblem} and \texttt{LinearVariationalSolver} objects. Defining the problem and solver objects allows the program to automatically update the primary field variable and thus inturn the discrete form of the system. Thus, we can solve the system inside a loop by simply calling the \texttt{solve()} method over the solver object. The call to \texttt{solve()} will create the stiffness matrix and the force vector and will not dispose of them immediately after execution, which allows the program to reuse them within the loop.
\begin{lstlisting}[language=Python,numbers=none,xleftmargin=0.0cm,backgroundcolor=\color{light-gray},frame=tlbr,framesep=4pt]
K = inner(density ** penal * sigma(u), grad(v)) * dx
problem= LinearVariationalProblem(K, F, u_new, bcs)
solver = LinearVariationalSolver(problem)
# FE-ANALYSIS --------------------------------------------------
solver_disp.solve()
\end{lstlisting}
FEniCS utilizes PETSc as its linear algebra backend and by default uses the sparse
LU decomposition ("lu") method, and the actual software that is called is then the sparse LU solver from UMFPACK (which PETSc has an interface to). We could easily change the solver parameters and design the solver as per our requirements by accessing the method parameters over the solver object.
\begin{lstlisting}[language=Python,numbers=none,xleftmargin=0.0cm,backgroundcolor=\color{light-gray},frame=tlbr,framesep=4pt]
solver.parameters["linear_solver"] = "gmres"
solver.parameters["preconditioner"] = "ilu"
\end{lstlisting}

\subsection{Optimality criteria based optimizer}
We initialize the density vector with \texttt{volfrac} as the initial density. Note that we use \texttt{as\char`_vector()} method over the density function to access the underlying array of values. The iterations start with the finite element analysis of the linear variational problem. The solution to the problem is stored in the variable \texttt{u\char`_sol}, which is then used to evaluate the objective function and sensitivity over each element of the mesh.
\begin{lstlisting}[language=Python,numbers=none,xleftmargin=0.0cm,backgroundcolor=\color{light-gray},frame=tlbr,framesep=4pt]
solver.solve()
objective = project(density ** penal * psi(u_sol), D).vector()[:]
sensitivity = -penal * (density.vector()[:]) ** (penal - 1) * project(psi(u_sol), D).vector()[:]
\end{lstlisting}
The optimality criteria based update of the density vector is done on lines 46 - 51. The bounds for the calculation of Lagrange multiplier by the bisection method are initialized on line 46 and the actual loop starts on line 47.
\begin{lstlisting}[language=Python,numbers=none,xleftmargin=0.0cm,backgroundcolor=\color{light-gray},frame=tlbr,framesep=4pt]
l1, l2, move = 0, 100000, 0.2
while l2 - l1 > 1e-4:
    l_mid = 0.5 * (l2 + l1)
    density_new = np.maximum(0.001,np.maximum(density.vector()[:] - move, np.minimum(1.0, np.minimum(density.vector()[:] + move, density.vector()[:] * np.sqrt(-sensitivity / l_mid)))))
    l1, l2 = (l_mid, l2) if sum(density_new) - volfrac * mesh.num_cells() > 0 else (l1, l_mid)
density.vector()[:] = density_new
\end{lstlisting}

\subsection{Mesh-independency filtering}
\label{sec:filter}
The modification of the sensitivities by the application of the sensitivity filter described by \cref{eq:filter} is a linear operation and this could be converted to a matrix product as defined by \cref{eq:filter_mat}. The matrix W is the weight matrix. The elements of the matrix satisfy the condition presented by \cref{eq:filter_cond} and \cref{fig:filter} . First we retrieve the midpoints of all the elements within the mesh and then evaluate the Euclidean distance matrix for the midpoint vector. In the case of topology optimization, suppose we have a collection of vectors $\{\bm{x}_i \in \mathbb{R}^d:i\in\{1,...,N\}\}$ containing the coordinates of midpoints ($\bm{x}_i$)of all the elements within the mesh. The Euclidean distance matrix is a $N \times N$ matrix that contains the pairwise distance between all the elements of a vector of length $N$, and is defined as:

\begin{equation}
\bm{D}_{i j}=\left(\mathbf{x}_{i}-\mathbf{x}_{j}\right)^{T}\left(\mathbf{x}_{i}-\mathbf{x}_{j}\right)=\left\|\mathbf{x}_{i}\right\|_{2}^{2}-2 \mathbf{x}_{i}^{T} \mathbf{x}_{j}+\left\|\mathbf{x}_{j}\right\|_{2}^{2}
\end{equation}

This gives us the pairwise distance between all the domain elements, or more specifically a single element $\bm{D}_{ab}$ represents the distance of $ele_a$ from $ele_b$. Then we apply the condition that the elements outside the zone of influence should have a weight equal to zero. i.e., the elements whose distance $>$ rmin should have a weight equal to zero.

\begin{remark}
By evaluating the weight matrix based on vector based calculations we reduced a lot of computational cost that is associated with evaluation of the same in a nested loop. The euclidean distance matrix is a symmetric matrix and thus there exists multiple algorithms to further reduce the computational cost associated with their evaluation on parallel system. The interested reader is referred to \citet{angelettiParallelEuclideanDistance2019, liChunkingMethodEuclidean2010} for the same.
\end{remark}
\begin{equation}
\label{eq:filter_mat}
        \left\{\begin{array}{c}
        \frac{\partial E_{1}}{\partial d_{1}} \\
        \vdots \\
        \frac{\partial E_{n}}{\partial d_{n}}
        \end{array}\right\}=
    \left(
        \left[\begin{array}{ccc}
        W_{11} & \cdots & W_{1 n} \\
        \vdots & \ddots & \vdots \\
        W_{n 1} & \cdots & W_{n n}
        \end{array}\right]
        \cdot
        \left(
            \left\{\begin{array}{c}
            d_{1} \\
            \vdots \\
            d_{n}
            \end{array}\right\} \odot\left\{\begin{array}{c}
            \frac{\partial E_{1}}{\partial d_{1}} \\
            \vdots \\
            \frac{\partial E_{n}}{\partial d_{n}}
            \end{array}\right\}
        \right)
    \right)
    \odot
    \left(
        \left\{\begin{array}{c}
        \frac{1}{d_{1}} \\
        \vdots \\
        \frac{1}{d_{n}}
        \end{array}\right\} \odot\left\{\begin{array}{c}
        \frac{1}{\sum_{b=1}^{n} W_{1 b}} \\
        \vdots \\
        \frac{1}{\sum_{b=1}^{n} W_{n b}}
        \end{array}\right\}
    \right)
\end{equation}
\begin{lstlisting}[language=Python,numbers=none,xleftmargin=0.0cm,backgroundcolor=\color{light-gray},frame=tlbr,framesep=4pt]
# PREPARE DISTANCE MATRICES FOR FILTER -----------------------------
midpoint = [cell.midpoint().array()[:] for cell in cells(mesh)]
distance_mat = rmin - sp.euclidean_distances(midpoint, midpoint)
distance_mat[distance_mat < 0] = 0
distance_sum = distance_mat.sum(1)  # sum the row
# FILTERING/MODIFICATION OF SENSITIVITIES ----------------------
sensitivity = np.divide(distance_mat @ np.multiply(density.vector()[:], sensitivity), np.multiply(density.vector()[:], distance_sum))
\end{lstlisting}
Once we have the weight matrix, we evaluate the vector of the sum of weight for each element. The modification in sensitivity is then evaluated by just performing the matrix operations described by \cref{eq:filter_mat}. Here $\odot$ represents the Hadamard product, and $(.)$ represents the dot product.  The distance matrix (weight matrix) and the distance sum vector do not vary during the iteration process, and thus they are evaluated and stored outside of the primary iteration loop.

    \section{Model Extensions}
\label{sec:extensions}

The original 99-line code by \citet{sigmund99LineTopology2001} and the 88-line code by \citet{andreassenEfficientTopologyOptimization2011}, the authors described how to extend their code to account for passive elements, other boundary conditions, and multiple load cases. In this section, we also describe these extensions based on our 55 lines of code. Besides, we show that the extension of the code to 3D problems still keeps the number of lines limited to 55 and requires a minimal change of the code.

\subsection{Working with 3D problems}
One of the amazing advantages of working with FEniCS is that it allows us to write a unified simulation code that can operate in $\mathbb{R}^d, d=1,2,3$, since the variational formulation is independent of dimension. By making a few modifications to the code we can use the same 55-lines of code to run 3D simulations. This is hugely beneficial to the previous implementation as to work with 3D they required significant modification of the code base \citep{andreassenEfficientTopologyOptimization2011}. To change the code into a 3D cantilever problem, we just need to update the mesh and add the third dimension values to the boundary condition and load definitions. Thus, only the following change is required:

\begin{lstlisting}[language=Python,numbers=none,xleftmargin=0.0cm,backgroundcolor=\color{light-gray},frame=tlbr,framesep=4pt]
def main(nelx, nely, nelz, volfrac, penal, rmin):
    mesh = BoxMesh(Point(0.0, 0.0, 0.0), Point(nelx, nely, nelz), nelx, nely, nelz)
    bcs = [DirichletBC(U, Constant((0.0, 0.0, 0.0)), support)]
    F = dot(v, Constant((0.0, -1.0, 0.0))) * ds(1)
\end{lstlisting}

\begin{figure}
    \centering
    \includegraphics[width=16cm]{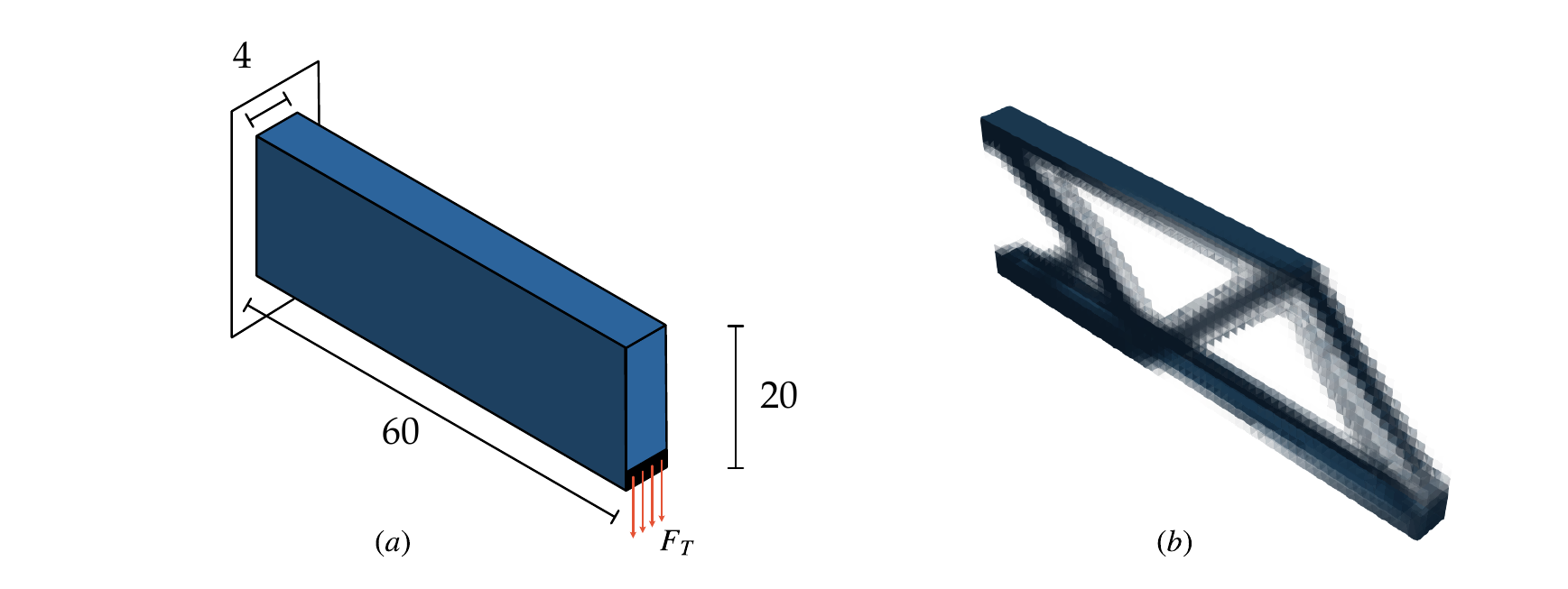}
    \caption{3D Cantilever beam problem. (a)The ground structure with the applied external loads and boundary conditions for the optimization of a 3D cantilever beam. (b) The topologically optimized design.}
    \label{fig:ext_3d}
\end{figure}

We need to add the number of element required in z-direction to the parameters of function definition. Changing the mesh constructor to \texttt{BoxMesh()}, creates a box partioned with (\texttt{nelx, nely, nelz}) elements. \cref{fig:ext_3d} shows the ground structure with the applied external loads and boundary conditions for the optimization of a 3D cantilever beam along with the optimized geometry achieved by running the command:

\begin{lstlisting}[language=Python,numbers=none,xleftmargin=0.0cm,backgroundcolor=\color{light-gray},frame=tlbr,framesep=4pt]
from topopt import main
main(nelx=60, nely=20, nelz=4, volfrac=0.3, penal=3.0, rmin=1.5)
\end{lstlisting}

\subsection{Other boundary conditions}
\begin{figure}
    \centering
    \includegraphics[width=16cm]{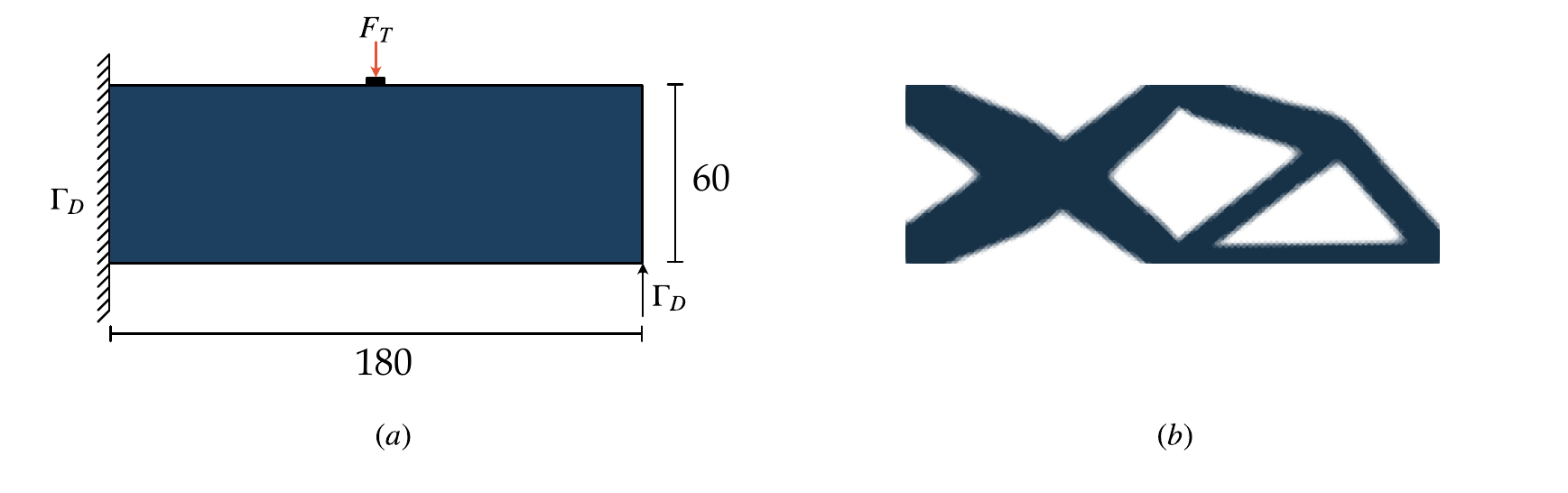}
    \caption{Propped cantilever beam problem. (a)The ground structure with the applied external loads and boundary conditions for the optimization of a propped cantilever beam. (b) Topologically optimized structure.}
    \label{fig:ext_bc}
\end{figure}

    The support elements are identified based on the conditional statements written in the C++ syntax. To add a prop to the other end of the cantilever beam we just need to identify the elements and then assign a Dirichlet boundary condition to it. This is achieved by adding the following lines of code:
\begin{lstlisting}[language=Python,numbers=none,xleftmargin=0.0cm,backgroundcolor=\color{light-gray},frame=tlbr,framesep=4pt]
prop_support = CompiledSubDomain("near(x[0], l,tol) && near(x[1], 0, tol)", tol=1e-14, l=nelx)
prop_bc = DirichletBC(U.sub(1), Constant(0.0), prop_support, method="pointwise")
bcs.append(prop_bc)
\end{lstlisting}
The solution presented in \cref{fig:ext_bc} is then obtained by the following:
\begin{lstlisting}[language=Python,numbers=none,xleftmargin=0.0cm,backgroundcolor=\color{light-gray},frame=tlbr,framesep=4pt]
from topopt import main
main(nelx=180, nely=60, volfrac=0.5, penal=3.0, rmin=3)
\end{lstlisting}

\subsection{Multiple load cases}
Adding multiple load cases to the problem is also achieved with less than 60 lines of code. In the two load case the change is made in line number 21 where we need to add a new marker for the second load. This is identified with the help of load marker 2. The information is then passed to the measure in a similar manner as before. We could then define loads on certain boundaries by using there specific markers.
 \begin{figure}
    \centering
    \includegraphics[width=16cm]{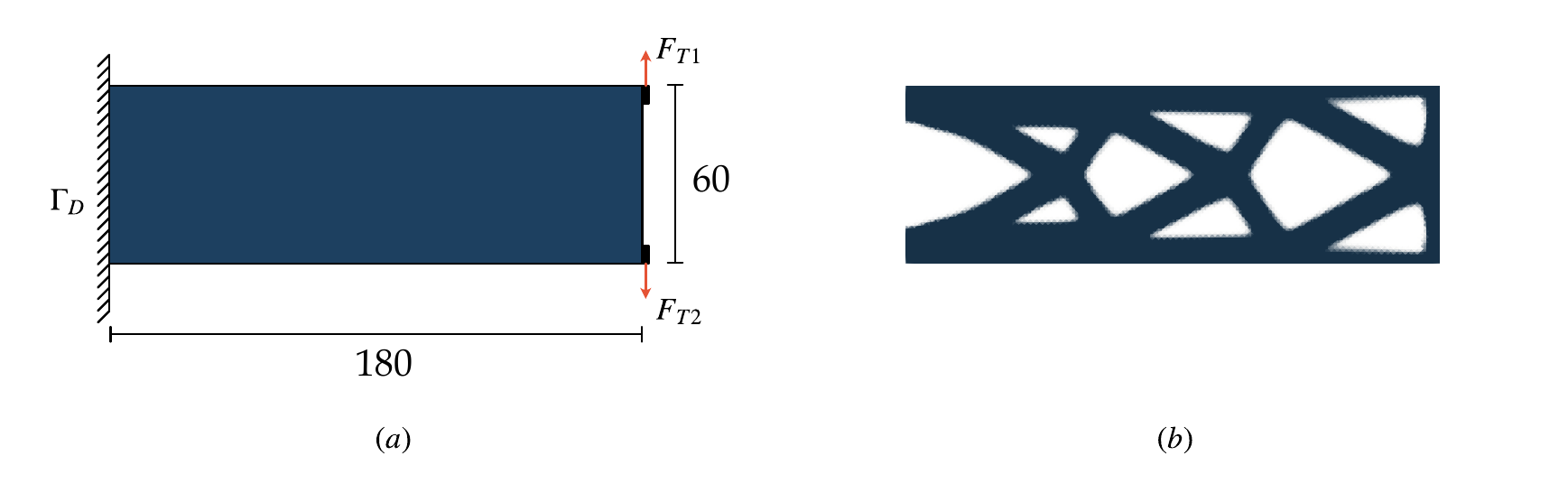}
    \caption{Cantilever beam with multiple load case. (a)The ground structure with the applied external loads and boundary conditions for the optimization of a cantilever beam with multiple load case. (b) Topologically optimized structure.}
    \label{fig:ext_lc}
\end{figure}
\begin{lstlisting}[language=Python,numbers=none,xleftmargin=0.0cm,backgroundcolor=\color{light-gray},frame=tlbr,framesep=4pt]
CompiledSubDomain("x[0]==l && x[1]<=2", l=nelx).mark(load_marker, 1)
CompiledSubDomain("x[0]==l && x[1]>=h-2", l=nelx, h=nely).mark(load_marker, 2)
ds = Measure("ds")(subdomain_data=load_marker)
F1 = dot(v, Constant((0.0, -1.0))) * ds(1)
F2 = dot(v, Constant((0.0, +1.0))) * ds(2)
problem = [LinearVariationalProblem(K, F1, u_sol, bcs),LinearVariationalProblem(K, F2, u_sol, bcs)]
\end{lstlisting}
The objective function and the sensitivity is calculated based on the displacement field by solving the two variational problems and summing them up for further calculations. This is done by replacing the lines 38-42 with the following.
\begin{lstlisting}[language=Python,numbers=none,xleftmargin=0.0cm,backgroundcolor=\color{light-gray},frame=tlbr,framesep=4pt]
objective, sensitivity = np.zeros(mesh.num_cells()), np.zeros(mesh.num_cells())
for i in range(2):
    # FE-ANALYSIS --------------------------------------------------
    solver = LinearVariationalSolver(problem[i])
    solver.solve()
    # OBJECTIVE FUNCTION AND SENSITIVITY ANALYSIS ------------------
    objective += project(density ** penal * psi(u_sol), D).vector()[:]
    sensitivity += -penal * (density.vector()[:]) ** (penal - 1) * project(psi(u_sol), D).vector()[:]
\end{lstlisting}

The solution presented in \cref{fig:ext_lc} is then obtained by the following:
\begin{lstlisting}[language=Python,numbers=none,xleftmargin=0.0cm,backgroundcolor=\color{light-gray},frame=tlbr,framesep=4pt]
from topopt import main
main(nelx=180, nely=60, volfrac=0.5, penal=3.0, rmin=3)
\end{lstlisting}

\subsection{Passive elements}
\begin{figure}
    \centering
    \includegraphics[width=16cm]{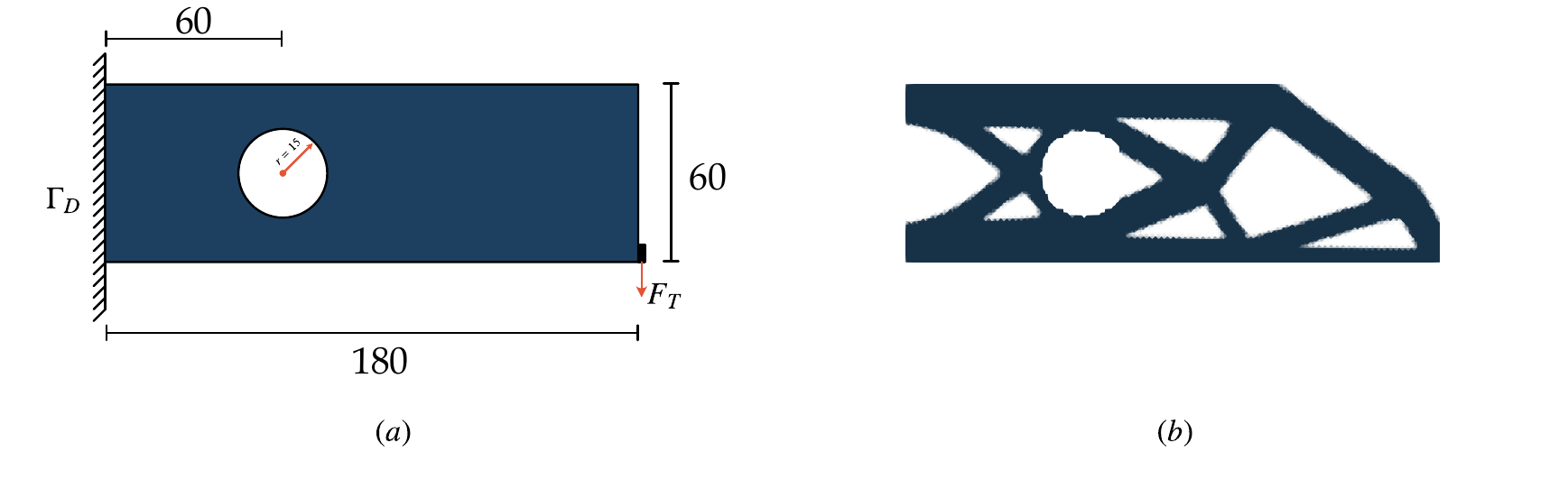}
    \caption{Cantilever beam with passive elements. (a)The ground structure with the applied external loads and boundary conditions for the optimization of a cantilever beam with a hole. (b) Topologically optimized structure.}
    \label{fig:ext_pe}
\end{figure}

To incorporate zones of fixed density, such as zones with no material (holes) or zones with fixed material, we define passive elements that have a fixed density throughout the iteration process. This is achieved with just additional five lines of code including the identification of the passive elements. The zone is identified and marked with the help of the Meshfunction class of FEniCS by adding the following lines of code.
\begin{lstlisting}[language=Python,numbers=none,xleftmargin=0.0cm,backgroundcolor=\color{light-gray},frame=tlbr,framesep=4pt]
# DEFINE PASSIVE ELEMENTS ------------------------------------------
circle = CompiledSubDomain('(x[0]-x0)*(x[0]-x0) + (x[1]-x1)*(x[1]-x1) <= r*r',x0=nelx/3, x1=nely/2, r=nely/4)
circle_marker = MeshFunction("size_t", mesh, mesh.topology().dim())
circle.mark(circle_marker, 1)
\end{lstlisting}
This information is then passed to the optimality criteria loop by adding the following line and then, explicitly setting their density value equal to minimum density in case of a hole.
\begin{lstlisting}[language=Python,numbers=none,xleftmargin=0.0cm,backgroundcolor=\color{light-gray},frame=tlbr,framesep=4pt]
density_new[circle_marker.where_equal(1)] = 0.001
\end{lstlisting}
The solution presented in \cref{fig:ext_pe} is then obtained by the following:
\begin{lstlisting}[language=Python,numbers=none,xleftmargin=0.0cm,backgroundcolor=\color{light-gray},frame=tlbr,framesep=4pt]
from topopt import main
main(nelx=180, nely=60, volfrac=0.5, penal=3.0, rmin=3)
\end{lstlisting}

\subsection{Alternative optimizers}
The optimality criteria based optimizer have been widely used in the literature because of its easy to understand formulation and straight forward implementation, but suffers from the lack of support for problems with multiple constrains. Another common optimizer that has been used successfully by many researches to solve multi-constraint topology optimization problems is based on the method of moving asymptotes(MMA) algorithm\citep{svanbergMethodMovingAsymptotes1987}. The MMA-algorithm is a mathematical programming algorithm, which is in nature similar to methods like Sequential Linear Programming (SLP) and Sequential Quadratic Programming (SQP) for solving smooth, non-linear optimization problems \citep{bendsoeTopologyOptimizationTheory2013}. The call to MMA requires definition of several input arguments along with the first and second derivatives of the objective as well as the constraint functions. In the case of TO the second order derivatives are assumed to be zero. One can find the definition of all the auxiliary variables in the documentation of code files which one can obtain from Prof. Krister Svanberg, KTH, Sweden. The MMA code is called by first defining the auxiliary variables as per the documentation and then replacing the optimality criteria block with the following:
\begin{lstlisting}[language=Python,numbers=none,xleftmargin=0.0cm,backgroundcolor=\color{light-gray},frame=tlbr,framesep=4pt]
xmma,ymma,zmma,lam,xsi,eta,aaa,zet,s,low,upp = \
        mmasub(m,n,loop,xval,xmin,xmax,xold1,xold2,f0val,df0dx,fval,dfdx,low,upp,a0,a,cMMa,d,move)
\end{lstlisting}

Since the code allows for direct access of the underlying matrices and vectors, one could also use the optimizers available in the open-source python libraries such a SciPy or PyOpt directly.

    \section{Extension to practical applications}
\begin{figure}
    \centering
    \includegraphics[width=15cm]{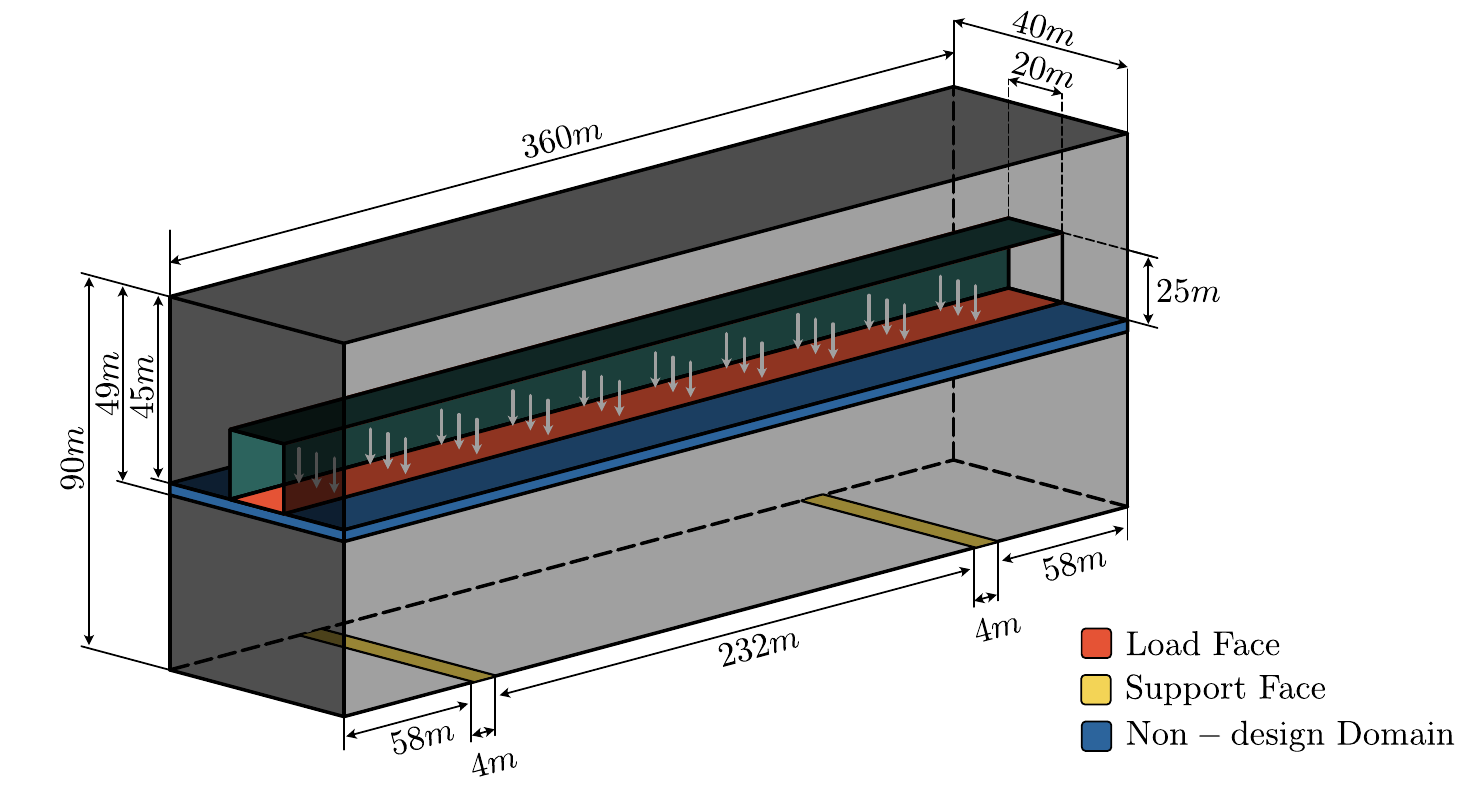}
    \caption{3D bridge problem. The ground structure is 360m long with an opening for traffic. It further consist of a non-designable deck and 4m wide support faces. The structure is meshed with 6 million elements.}
    \label{fig:bridge_3d}
\end{figure}

This code is capable of running large scale simulations on systems ranging from high performance workstations to computational clusters. The readability of the code allows for easy modifications and the just in time compilation provides us with speed and efficiency. With a few modifications, the code can handle complex structural configuration with support for parallel computation. In this section we will discuss few challenges related to application of TO to large scale engineering problems with the help of a 3D bridge example and its implementation with the 55-line code. \cref{fig:bridge_3d} shows the ground structure with the applied external loads and boundary conditions for the optimization of a 3D bridge with a non-designable deck. An opening of size $20m \times 25m$ is placed over the non-designable deck and a uniformly distributed load is applied over the deck inside the opening. The structure is meshed with approximately six-million tetrahedral elements with three-million DoFs.

 \begin{figure}
    \centering
    \includegraphics[width=11cm]{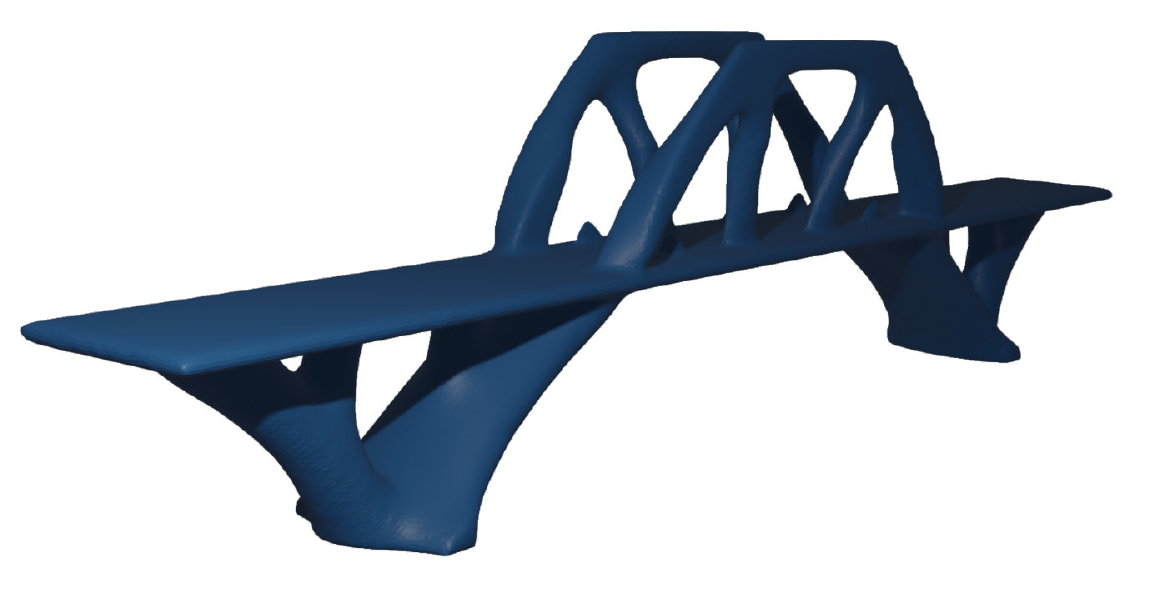}
    \caption{3D bridge problem. Smoothened topologically optimized design.}
    \label{fig:bridge_3d_result}
\end{figure}
% I will discuss the three sections with the same base problem of arch bridge.
\subsection{Complex structural configuration}
The ground structure for the 3D-bridge as presented in \cref{fig:bridge_3d} consists of three sub-domains: the non-designable deck, the load face and the support face. Such kind of structural configuration is difficult to handle in the MATLAB codes \citep{sigmund99LineTopology2001,andreassenEfficientTopologyOptimization2011,ferrariNewGeneration992020} but are comparatively easy to handle in commercial packages in which we can have an explicit representation of the complex structural configuration \citep{zuoSimpleCompactPython2015}. FEniCS supports marked computational mesh generated by 'gmsh'\citep{gmsh}; which allows us to explicitly mark different sub-domains in 'gmsh' with numerical labels and then  read those directly into FEniCS. This extends the capability of the 55-line code to complex structural configurations. For complex geometries one needs to follow the mesh processing pipeline, and, thus create the mesh in 'gmsh' and mark different sub-domains, convert the mesh to XDMF with 'meshio'\citep{nicoschlomer2018} and then finally read the mesh directly into the program with the commands:

\begin{lstlisting}[language=Python,numbers=none,xleftmargin=0.0cm,backgroundcolor=\color{light-gray},frame=tlbr,framesep=4pt]
mesh = Mesh()
with XDMFFile("domain.xdmf") as infile:
    infile.read(mesh)
\end{lstlisting}

The supports and loads are defined by reading the numerical labels of the mesh and defining 'Measures'\texttt{(dx, ds)} over the domain based on these labels. This is achieved by replacing lines $17-22$ with the following:
\begin{lstlisting}[language=Python,numbers=none,xleftmargin=0.0cm,backgroundcolor=\color{light-gray},frame=tlbr,framesep=4pt]
mvc = MeshValueCollection("size_t", mesh, 2)
with XDMFFile("mesh/surface.xdmf") as infile:
        infile.read(mvc, "surface_tag")
mf = cpp.mesh.MeshFunctionSizet(mesh, mvc)
ds = Measure("ds")(subdomain_data=mf)
bcs = [DirichletBC(U, Constant((0.0, 0.0, 0.0)), mf, support_tag)]
F = dot(v, Constant((0.0, -1.0, 0.0))) * ds(load_tag)
\end{lstlisting}

The imported mesh will work directly with code as the computation of the filter is based on the actual centre-points of the elements (refer \cref{sec:filter}) as opposed to the assumptions made in the 88 or 99 line codes where the filter is evaluated based on the assumption of uniform mesh with unit cells. The optimized geometry presented in \cref{fig:bridge_3d_result} is achieved by running the command:

\begin{lstlisting}[language=Python,numbers=none,xleftmargin=0.0cm,backgroundcolor=\color{light-gray},frame=tlbr,framesep=4pt]
from topopt import main
main(volfrac=0.15, penal=3.0, rmin=10.0)
\end{lstlisting}

\subsection{Handling large 3D problems}

Solving large scale 3D problems require access to systems with large amount of RAM and computational power. Even with powerful computational systems the implementation could be slow because of inefficient coding. The computational cost associated with TO can be attributed to three major operations: computation of the filter, evaluation of the displacement solution of the problem, and evaluation of the optimal solution of the optimization problem. Since all of these are dependent on the size of the mesh, the computational cost of the algorithm is directly related to the mesh.  

\begin{figure}
    \centering
    \includegraphics[width=16cm]{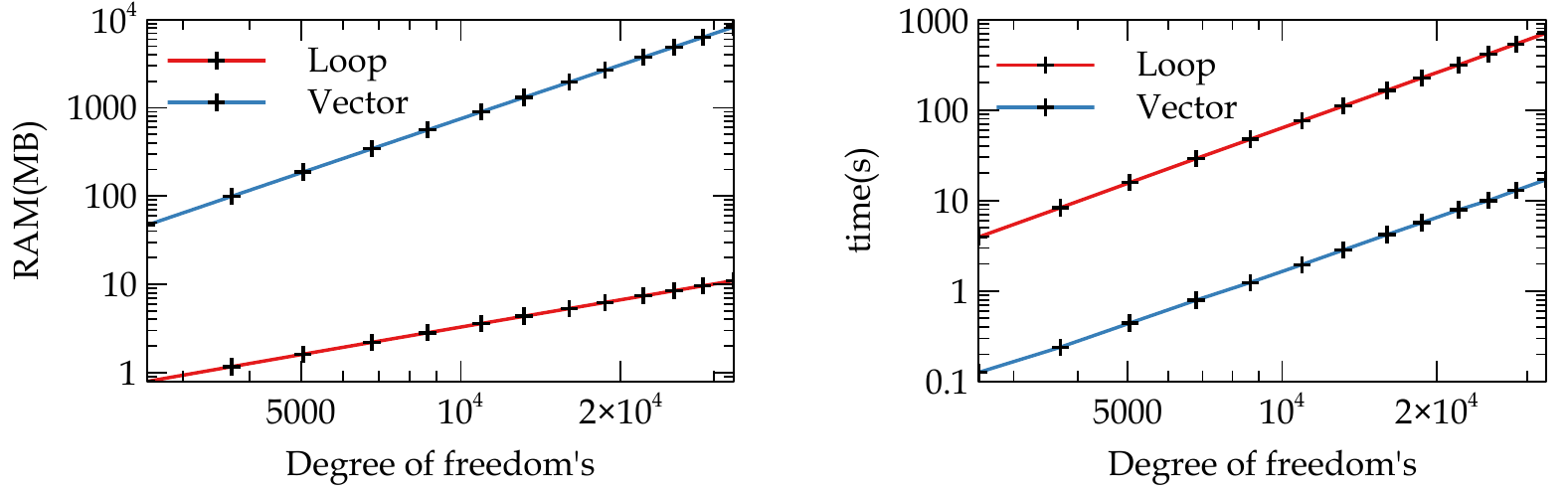}
    \caption{Comparison of looping versus vector operation for computation of filter. With increasing number of degrees of freedom's looping approach takes substantially less amount of RAM as compared to vector approach but require more time.}
    \label{fig:loop_vec}
\end{figure}

In the original 99-line code and further it's evolution into the 88-line code, loops were used to evaluate the weight-matrices for the filter. Loops are known to be extremely slow as compared to vector operations, but, vector operations require a RAM size capable of storing the full matrices required to perform the computation. In \cref{fig:loop_vec}, we can see that in terms of RAM, loops perform much better as compared to the vector approach, but, in terms of computational time the vector approach outperforms loop. For 33000 DoF's the RAM required by the looping approach is a mere 11MB as compared to 8GB required by vector approach, but, vector approach completes the computation in 17 seconds as opposed to 710 seconds taken by loops. To take benefit of both the approaches it is advised to use the chunking technique as described in \citet{liChunkingMethodEuclidean2010} by looping over large chunks of the mid-point vector which could be stored in the RAM for computation. 

In the 88-line MATLAB code \citep{andreassenEfficientTopologyOptimization2011} substantial reduction in computational expense as compared to the original 99-line code \citep{sigmund99LineTopology2001} was achieved by vectorization of the loops and the new 99-line code \citep{ferrariNewGeneration992020} further provides a reduction in computational expense by smart use of sparse matrices and using the 'fsparse' function from the \texttt{stenglib} library that is written in 'C'. The use of sparse solvers along with vectorization has lead to substantial reduction in the computational expense of the evaluation of the displacement solution of the problem and also evaluation of the optimal solution of the optimization problem. The latest 99-line code achieves the assembly of the stiffness matrix of a six million DoF's system in under 10 seconds. Since we are using PetSc as the backend, our code benefits from the capabilities of the library such as support for parallelization, sparse matrices and iterative solvers. This results in under 10 second assembly of stiffness matrix for the same six million DoF's system, and with parallelization we can achieve further speedups.
  
\subsection{Parallel implementation of topology optimization}
Topology optimization algorithm requires computation of both the displacement solution of the systems and the optimization solution of the system in each iteration. To solve large scale engineering structures researchers in the past have parallelized the TO algorithm \citep{aageParallelFrameworkTopology2013b,borrvallLargescaleTopologyOptimization2001,challisHighResolutionTopology2014,evgrafovLargescaleParallelTopology2008a,schmidt2589LineTopology2011b,vemagantiParallelMethodsOptimality2005b,wadbroMegapixelTopologyOptimization2009,aageTopologyOptimizationUsing2015a,martinez-frutosEfficientTopologyOptimization2017, baigesLargescaleStochasticTopology2019,martinez-frutosGPUAccelerationEvolutionary2017,martinez-frutosLargescaleRobustTopology2016}. Many of these works are extremely capable in solving the TO problem efficiently on parallel HPC systems, but are too complex for beginners to understand and implement. In this section our aim is to present the parallel performance of our code and help beginners take benefit of multi-core simulations on there laptops or HPC workstations. This implementation by no means is as optimized as the  2589-line code by \citet{schmidt2589LineTopology2011b} or the 6300-lines code of \citet{aageTopologyOptimizationUsing2015a}, but serves as a proof-of-concept for the parallel capability of the 55-line code.

The modifications necessary to make the code compatible for parallel computation lies only in the modification of filter. Mesh-independency filter as described in \cref{sec:mif} requires information of the co-ordinates of the center-points of elements in a zone around the element under consideration. Since, parallelization in FEniCS works by splitting the mesh into 'n' parts - where 'n' is the number of processors - and then passing only a portion of the mesh to a single processor, the information required for computation of the filter is not available with each processor. Thus, to make the code compatible for parallel computation we have to change the filter to the Helmholtz filter as described in \citep{lazarovFiltersTopologyOptimization2011, desouzaTopologyOptimizationApplied2020}. Parallel computation can then be performed on 16 cores by running the command:
\begin{lstlisting}[language=Python,numbers=none,xleftmargin=0.0cm,backgroundcolor=\color{light-gray},frame=tlbr,framesep=4pt]
mpirun -np 16 python3 main.py
\end{lstlisting}
In the above command \texttt{main.py} file contains the following code:
\begin{lstlisting}[language=Python,numbers=none,xleftmargin=0.0cm,backgroundcolor=\color{light-gray},frame=tlbr,framesep=4pt]
from topopt import main
main(volfrac=0.15, penal=3.0, rmin=10.0)
\end{lstlisting}

It is to be noted that, our implementation supports complex structural configuration on parallel systems. For the problem presented in \cref{fig:bridge_3d} with six million elements, the code took 10 minutes and 50 seconds per iteration on a single core, and, 1 minute and 40 seconds per iteration on 16 cores. This proves the code's scalability, and with further improvements in solver design and the use of better optimization algorithms, the code is capable of substantial computational gains.

    \section{Conclusion}
\label{sec:conclusion}
This paper provides a simple, compact, and easy-to-understand implementation of topology optimization using the solid isotropic material with penalization methodology for 2D and 3D structures via an open-source finite element python package (FEniCS). By revisiting the original 99-line code by \citet{sigmund99LineTopology2001} and its further modification to 88-lines by \citet{andreassenEfficientTopologyOptimization2011} and the new 99-line by \citet{ferrariNewGeneration992020}, we present a brief step-by-step overview of the topology optimization problem's mathematical model and a detailed description of the FEniCS implementation.

The 99-line code laid the foundation and has helped many researchers enter the field of topology optimization, emphasizing the 2D plane stress problem. The 88-line code significantly reduced the computational time by vectorizing the loops, while also reducing the number of lines to 88. The new-99-line code further improved the efficiency of the original code by updating the code to latest developments in the sparse computation of matrices in MATLAB. This implementation's key strength is the generalized and compact approach enabled by FEniCS, which allows us to make a unified code-base for 2D and 3D problems with easy to understand UFL expressions that make the code extremely readable in just 55-lines. 

Compactness in the code is also achieved by vectorizing most of the operations. By utilizing the Euclidean distance matrix concept, we have also vectorized the weight matrix computation for mesh-independency filters which has lead to substantial savings in the computational time for evaluation of the filter. This also allows us to evaluate the filter on complex structural configurations without any change in the code-base. We also demonstrate the capability of the code to handle complex large scale engineering structures with support for parallel computation on high performance workstations.

Since the expressions written in the python code are automatically converted into compiled C++ functions, the implementation is computationally efficient. The whole code-base is based on open-source packages, and thus, it is accessible and ready-to-use by the entire research community. Considering FEniCS is platform-independent \citep{haleContainersPortableProductive2017}, the code works on Windows, MAC and Linux systems alike. FEniCS is also equipped with powerful tools that allow straightforward extension of the current code into different areas, such as nonlinearity, hyper-elasticity, compliant mechanisms, and multi-scale modeling with the same base code.
\section{Declarations}
\subsection{Funding}
The authors gratefully acknowledge financial support from the Ministry of Human Resource Development. The second author (Rajib Chowdhury) thanks the funding support from the SERB via file no. CRG/2019/004600 and DRDL via file no. DRDL/24/08P/19/0235/43386.

\subsection{Conflict of interest}
The authors declare that they have no known competing financial or personal relationship with other people or organizations that could inappropriately influence or bias the content of the paper.
%\subsection{Availability of data and material}
%The datasets generated during and/or analyzed during the current study are available in the \textbf{topo-fenics} repository, \url{https://github.com/iitrabhi/topo-fenics}
%\subsection{Code availability}
%The code used for the current study is available in the \textbf{topo-fenics} repository, \url{https://github.com/iitrabhi/topo-fenics}
%\subsection{Authors' contributions}
%A. Gupta was responsible for the study conception and design. The first draft of the manuscript was written by A. Gupta and all authors commented on previous versions of the manuscript. All authors read and approved the final manuscript.
%\subsection{Replication of results}
%To facilitate replication of the results of this paper, the source code is presented in \ref{sec:appenix_a}.

    \appendix
%\crefalias{section}{appsec}
\section{FEniCS Code}\label{sec:appenix_a}
\begin{lstlisting}[language=Python
]
from dolfin import *
import numpy as np, sklearn.metrics.pairwise as sp
# A 55 LINE TOPOLOGY OPTIMIZATION CODE ---------------------------------
def main(nelx, nely, volfrac, penal, rmin):
    sigma = lambda _u: 2.0 * mu * sym(grad(_u)) + lmbda * tr(sym(grad(_u))) * Identity(len(_u))
    psi = lambda _u: lmbda / 2 * (tr(sym(grad(_u))) ** 2) + mu * tr(sym(grad(_u)) * sym(grad(_u)))
    xdmf = XDMFFile("output/density.xdmf")
    mu, lmbda = Constant(0.4), Constant(0.6)
    # PREPARE FINITE ELEMENT ANALYSIS ----------------------------------
    mesh = RectangleMesh(Point(0, 0), Point(nelx, nely), nelx, nely, "right/left")
    U = VectorFunctionSpace(mesh, "P", 1)
    D = FunctionSpace(mesh, "DG", 0)
    u, v = TrialFunction(U), TestFunction(U)
    u_sol, density_old, density = Function(U), Function(D), Function(D, name="density")
    density.vector()[:] = volfrac
    # DEFINE SUPPORT ---------------------------------------------------
    support = CompiledSubDomain("near(x[0], 0.0, tol) && on_boundary", tol=1e-14)
    bcs = [DirichletBC(U, Constant((0.0, 0.0)), support)]
    # DEFINE LOAD ------------------------------------------------------
    load_marker = MeshFunction("size_t", mesh, mesh.topology().dim() - 1)
    CompiledSubDomain("x[0]==l && x[1]<=2", l=nelx).mark(load_marker, 1)
    ds = Measure("ds")(subdomain_data=load_marker)
    F = dot(v, Constant((0.0, -1.0))) * ds(1)
    # SET UP THE VARIATIONAL PROBLEM AND SOLVER ------------------------
    K = inner(density ** penal * sigma(u), grad(v)) * dx
    problem = LinearVariationalProblem(K, F, u_sol, bcs)
    solver = LinearVariationalSolver(problem)
    # PREPARE DISTANCE MATRICES FOR FILTER -----------------------------
    midpoint = [cell.midpoint().array()[:] for cell in cells(mesh)]
    distance_mat = rmin - sp.euclidean_distances(midpoint, midpoint)
    distance_mat[distance_mat < 0] = 0
    distance_sum = distance_mat.sum(1)
    # START ITERATION --------------------------------------------------
    loop, change = 0, 1
    while change > 0.01 and loop < 2000:
        loop = loop + 1
        density_old.assign(density)
        # FE-ANALYSIS --------------------------------------------------
        solver.solve()
        # OBJECTIVE FUNCTION AND SENSITIVITY ANALYSIS ------------------
        objective = project(density ** penal * psi(u_sol), D).vector()[:]
        sensitivity = -penal * (density.vector()[:]) ** (penal - 1) * project(psi(u_sol), D).vector()[:]
        # FILTERING/MODIFICATION OF SENSITIVITIES ----------------------
        sensitivity = np.divide(distance_mat @ np.multiply(density.vector()[:], sensitivity), np.multiply(density.vector()[:], distance_sum))
        # DESIGN UPDATE BY THE OPTIMALITY CRITERIA METHOD --------------
        l1, l2, move = 0, 100000, 0.2
        while l2 - l1 > 1e-4:
            l_mid = 0.5 * (l2 + l1)
            density_new = np.maximum(0.001,np.maximum(density.vector()[:] - move, np.minimum(1.0, np.minimum(density.vector()[:] + move, density.vector()[:] * np.sqrt(-sensitivity / l_mid)))))
            l1, l2 = (l_mid, l2) if sum(density_new) - volfrac * mesh.num_cells() > 0 else (l1, l_mid)
        density.vector()[:] = density_new
        # PRINT RESULTS ------------------------------------------------
        change = norm(density.vector()-density_old.vector(), norm_type="linf", mesh=mesh)
        print("it.: {0} , obj.: {1:.3f} Vol.: {2:.3f}, ch.: {3:.3f}".format(loop, sum(objective), sum(density.vector()[:]) / mesh.num_cells(), change))
        xdmf.write(density, loop)
\end{lstlisting}

    %% \nocite{*}
    %\section*{References}
\bibliography{topology-fenics}

\begin{thebibliography}{77}
\providecommand{\natexlab}[1]{#1}
\providecommand{\url}[1]{\texttt{#1}}
\expandafter\ifx\csname urlstyle\endcsname\relax
  \providecommand{\doi}[1]{doi: #1}\else
  \providecommand{\doi}{doi: \begingroup \urlstyle{rm}\Url}\fi

\bibitem[Aage and
  Lazarov(2013{\natexlab{a}})]{aageParallelFrameworkTopology2013}
N.~Aage and B.~S. Lazarov.
\newblock Parallel framework for topology optimization using the method of
  moving asymptotes.
\newblock \emph{Structural and multidisciplinary optimization}, 47\penalty0
  (4):\penalty0 493--505, 2013{\natexlab{a}}.

\bibitem[Aage and
  Lazarov(2013{\natexlab{b}})]{aageParallelFrameworkTopology2013b}
N.~Aage and B.~S. Lazarov.
\newblock Parallel framework for topology optimization using the method of
  moving asymptotes.
\newblock \emph{Structural and Multidisciplinary Optimization}, 47\penalty0
  (4):\penalty0 493--505, Apr. 2013{\natexlab{b}}.
\newblock ISSN 1615-1488.
\newblock \doi{10.1007/s00158-012-0869-2}.

\bibitem[Aage et~al.(2015{\natexlab{a}})Aage, Andreassen, and
  Lazarov]{aageTopologyOptimizationUsing2015}
N.~Aage, E.~Andreassen, and B.~S. Lazarov.
\newblock Topology optimization using {{PETSc}}: {{An}} easy-to-use, fully
  parallel, open source topology optimization framework.
\newblock \emph{Structural and Multidisciplinary Optimization}, 51\penalty0
  (3):\penalty0 565--572, Mar. 2015{\natexlab{a}}.
\newblock \doi{10.1007/s00158-014-1157-0}.

\bibitem[Aage et~al.(2015{\natexlab{b}})Aage, Andreassen, and
  Lazarov]{aageTopologyOptimizationUsing2015a}
N.~Aage, E.~Andreassen, and B.~S. Lazarov.
\newblock Topology optimization using {{PETSc}}: {{An}} easy-to-use, fully
  parallel, open source topology optimization framework.
\newblock \emph{Structural and Multidisciplinary Optimization}, 51\penalty0
  (3):\penalty0 565--572, Mar. 2015{\natexlab{b}}.
\newblock ISSN 1615-1488.
\newblock \doi{10.1007/s00158-014-1157-0}.

\bibitem[Aage et~al.(2017)Aage, Andreassen, Lazarov, and
  Sigmund]{aageGigavoxelComputationalMorphogenesis2017a}
N.~Aage, E.~Andreassen, B.~S. Lazarov, and O.~Sigmund.
\newblock Giga-voxel computational morphogenesis for structural design.
\newblock \emph{Nature}, 550\penalty0 (7674):\penalty0 84--86, 2017.

\bibitem[Abhyankar et~al.(2018)Abhyankar, Brown, Constantinescu, Ghosh, Smith,
  and Zhang]{abhyankar2018petsc}
S.~Abhyankar, J.~Brown, E.~M. Constantinescu, D.~Ghosh, B.~F. Smith, and
  H.~Zhang.
\newblock Petsc/ts: A modern scalable ode/dae solver library.
\newblock \emph{arXiv preprint arXiv:1806.01437}, 2018.

\bibitem[Ahrens(2005)]{paraview}
G.~B. L.~C. Ahrens, James.
\newblock \emph{ParaView: An End-User Tool for Large Data Visualization,
  Visualization Handbook}, Aug. 2005.

\bibitem[Aln\ae{}s et~al.(2009)Aln\ae{}s, Logg, Mardal, Skavhaug, and
  Langtangen]{UFCAlnaesLoggEtAl2009a}
M.~S. Aln\ae{}s, A.~Logg, K.-A. Mardal, O.~Skavhaug, and H.~P. Langtangen.
\newblock Unified framework for finite element assembly.
\newblock \emph{International Journal of Computational Science and
  Engineering}, 4\penalty0 (4):\penalty0 231--244, 2009.
\newblock \doi{10.1504/IJCSE.2009.029160}.

\bibitem[Aln\ae{}s et~al.(2014)Aln\ae{}s, Logg, \O{}lgaard, Rognes, and
  Wells]{UFLAlnaesEtAl2012}
M.~S. Aln\ae{}s, A.~Logg, K.~B. \O{}lgaard, M.~E. Rognes, and G.~N. Wells.
\newblock Unified form language: A domain-specific language for weak
  formulations of partial differential equations.
\newblock \emph{ACM Transactions on Mathematical Software}, 40\penalty0 (2),
  2014.
\newblock \doi{10.1145/2566630}.

\bibitem[Aln{\ae}s et~al.(2015)Aln{\ae}s, Blechta, Hake, Johansson, Kehlet,
  Logg, Richardson, Ring, Rognes, and Wells]{FEniCSAlnaesBlechta2015a}
M.~S. Aln{\ae}s, J.~Blechta, J.~Hake, A.~Johansson, B.~Kehlet, A.~Logg,
  C.~Richardson, J.~Ring, M.~E. Rognes, and G.~N. Wells.
\newblock The fenics project version 1.5.
\newblock \emph{Archive of Numerical Software}, 3\penalty0 (100), 2015.
\newblock \doi{10.11588/ans.2015.100.20553}.

\bibitem[Amir et~al.(2014)Amir, Aage, and
  Lazarov]{amirMultigridCGEfficientTopology2014}
O.~Amir, N.~Aage, and B.~S. Lazarov.
\newblock On multigrid-{{CG}} for efficient topology optimization.
\newblock \emph{Structural and Multidisciplinary Optimization}, 49\penalty0
  (5):\penalty0 815--829, 2014.

\bibitem[Andreassen et~al.(2011)Andreassen, Clausen, Schevenels, Lazarov, and
  Sigmund]{andreassenEfficientTopologyOptimization2011}
E.~Andreassen, A.~Clausen, M.~Schevenels, B.~S. Lazarov, and O.~Sigmund.
\newblock Efficient topology optimization in {{MATLAB}} using 88 lines of code.
\newblock \emph{Structural and Multidisciplinary Optimization}, 43\penalty0
  (1):\penalty0 1--16, Jan. 2011.
\newblock \doi{10.1007/s00158-010-0594-7}.

\bibitem[Angeletti et~al.(2019)Angeletti, Bonny, and
  Koko]{angelettiParallelEuclideanDistance2019}
M.~Angeletti, J.-M. Bonny, and J.~Koko.
\newblock Parallel {{Euclidean}} distance matrix computation on big datasets *.
\newblock Feb. 2019.

\bibitem[Baiges et~al.(2019)Baiges, {Mart{\'i}nez-Frutos}, {Herrero-P{\'e}rez},
  Otero, and Ferrer]{baigesLargescaleStochasticTopology2019}
J.~Baiges, J.~{Mart{\'i}nez-Frutos}, D.~{Herrero-P{\'e}rez}, F.~Otero, and
  A.~Ferrer.
\newblock Large-scale stochastic topology optimization using adaptive mesh
  refinement and coarsening through a two-level parallelization scheme.
\newblock \emph{Computer Methods in Applied Mechanics and Engineering},
  343:\penalty0 186--206, 2019.

\bibitem[Balay et~al.(1997)Balay, Gropp, McInnes, and Smith]{petsc-efficient}
S.~Balay, W.~D. Gropp, L.~C. McInnes, and B.~F. Smith.
\newblock Efficient management of parallelism in object oriented numerical
  software libraries.
\newblock In E.~Arge, A.~M. Bruaset, and H.~P. Langtangen, editors,
  \emph{Modern Software Tools in Scientific Computing}, pages 163--202.
  Birkh{\"{a}}user Press, 1997.

\bibitem[Balay et~al.(2010)Balay, Buschelman, Eijkhout, Gropp, Kaushik,
  Knepley, Mcinnes, Smith, and Zhang]{PetscBalay2010}
S.~Balay, K.~Buschelman, V.~Eijkhout, W.~Gropp, D.~Kaushik, M.~Knepley, L.~C.
  Mcinnes, B.~Smith, and H.~Zhang.
\newblock {PETSc Users Manual}.
\newblock \emph{ReVision}, \penalty0 (ANL-95/11 - Revision 3.12), 2010.
\newblock \doi{10.2172/1178104}.
\newblock URL \url{https://www.mcs.anl.gov/petsc}.

\bibitem[Balay et~al.(2019)Balay, Abhyankar, Adams, Brown, Brune, Buschelman,
  Dalcin, Dener, Eijkhout, Gropp, Karpeyev, Kaushik, Knepley, May, McInnes,
  Mills, Munson, Rupp, Sanan, Smith, Zampini, Zhang, and Zhang]{petsc-web-page}
S.~Balay, S.~Abhyankar, M.~F. Adams, J.~Brown, P.~Brune, K.~Buschelman,
  L.~Dalcin, A.~Dener, V.~Eijkhout, W.~D. Gropp, D.~Karpeyev, D.~Kaushik, M.~G.
  Knepley, D.~A. May, L.~C. McInnes, R.~T. Mills, T.~Munson, K.~Rupp, P.~Sanan,
  B.~F. Smith, S.~Zampini, H.~Zhang, and H.~Zhang.
\newblock {PETS}c {W}eb page.
\newblock \url{https://www.mcs.anl.gov/petsc}, 2019.
\newblock URL \url{https://www.mcs.anl.gov/petsc}.

\bibitem[Bends{\o}e(1989)]{bendsoeOptimalShapeDesign1989}
M.~P. Bends{\o}e.
\newblock Optimal shape design as a material distribution problem.
\newblock \emph{Structural optimization}, 1\penalty0 (4):\penalty0 193--202,
  Dec. 1989.
\newblock \doi{10.1007/BF01650949}.

\bibitem[Bends{\o}e(1995)]{bendsoeOptimizationStructuralTopology1995}
M.~P. Bends{\o}e.
\newblock \emph{Optimization of {{Structural Topology}}, {{Shape}}, and
  {{Material}}}.
\newblock {Springer Berlin Heidelberg}, {Berlin, Heidelberg}, 1995.
\newblock ISBN 978-3-662-03117-9 978-3-662-03115-5.
\newblock \doi{10.1007/978-3-662-03115-5}.

\bibitem[Bends{\o}e and Kikuchi(1988)]{bendsoeGeneratingOptimalTopologies1988}
M.~P. Bends{\o}e and N.~Kikuchi.
\newblock Generating optimal topologies in structural design using a
  homogenization method.
\newblock \emph{Computer Methods in Applied Mechanics and Engineering},
  71\penalty0 (2):\penalty0 197--224, Nov. 1988.
\newblock \doi{10.1016/0045-7825(88)90086-2}.

\bibitem[Bendsoe and Sigmund(2013)]{bendsoeTopologyOptimizationTheory2013}
M.~P. Bendsoe and O.~Sigmund.
\newblock \emph{Topology Optimization: Theory, Methods, and Applications}.
\newblock {Springer Science \& Business Media}, 2013.

\bibitem[Bleyer(2018)]{bleyer2018numericaltours}
J.~Bleyer.
\newblock Numerical tours of computational mechanics with {FE}ni{CS}.
\newblock 2018.
\newblock \doi{10.5281/zenodo.1287832}.

\bibitem[Borrvall and
  Petersson(2001)]{borrvallLargescaleTopologyOptimization2001}
T.~Borrvall and J.~Petersson.
\newblock Large-scale topology optimization in {{3D}} using parallel computing.
\newblock \emph{Computer Methods in Applied Mechanics and Engineering},
  190\penalty0 (46):\penalty0 6201--6229, Sept. 2001.
\newblock ISSN 0045-7825.
\newblock \doi{10.1016/S0045-7825(01)00216-X}.

\bibitem[Challis(2010)]{challisDiscreteLevelsetTopology2010}
V.~J. Challis.
\newblock A discrete level-set topology optimization code written in
  {{Matlab}}.
\newblock \emph{Structural and Multidisciplinary Optimization}, 41\penalty0
  (3):\penalty0 453--464, Apr. 2010.
\newblock \doi{10.1007/s00158-009-0430-0}.

\bibitem[Challis et~al.(2014)Challis, Roberts, and
  Grotowski]{challisHighResolutionTopology2014}
V.~J. Challis, A.~P. Roberts, and J.~F. Grotowski.
\newblock High resolution topology optimization using graphics processing units
  ({{GPUs}}).
\newblock \emph{Structural and Multidisciplinary Optimization}, 49\penalty0
  (2):\penalty0 315--325, Feb. 2014.
\newblock ISSN 1615-1488.
\newblock \doi{10.1007/s00158-013-0980-z}.

\bibitem[Chen et~al.(2019)Chen, Zhang, and
  Zhu]{chen213lineTopologyOptimization2019}
Q.~Chen, X.~Zhang, and B.~Zhu.
\newblock A 213-line topology optimization code for geometrically nonlinear
  structures.
\newblock \emph{Structural and Multidisciplinary Optimization}, 59\penalty0
  (5):\penalty0 1863--1879, May 2019.
\newblock \doi{10.1007/s00158-018-2138-5}.

\bibitem[{de Souza} and Silva(2020)]{desouzaTopologyOptimizationApplied2020}
E.~M. {de Souza} and E.~C.~N. Silva.
\newblock Topology optimization applied to the design of actuators driven by
  pressure loads.
\newblock \emph{Structural and Multidisciplinary Optimization}, 61\penalty0
  (5):\penalty0 1763--1786, May 2020.
\newblock ISSN 1615-147X, 1615-1488.
\newblock \doi{10.1007/s00158-019-02421-5}.

\bibitem[Evgrafov et~al.(2008)Evgrafov, Rupp, Maute, and
  Dunn]{evgrafovLargescaleParallelTopology2008a}
A.~Evgrafov, C.~J. Rupp, K.~Maute, and M.~L. Dunn.
\newblock Large-scale parallel topology optimization using a dual-primal
  substructuring solver.
\newblock \emph{Structural and Multidisciplinary Optimization}, 36\penalty0
  (4):\penalty0 329--345, Oct. 2008.
\newblock ISSN 1615-1488.
\newblock \doi{10.1007/s00158-007-0190-7}.

\bibitem[Ferrari and Sigmund(2020)]{ferrariNewGeneration992020}
F.~Ferrari and O.~Sigmund.
\newblock A new generation 99 line {{Matlab}} code for compliance {{Topology
  Optimization}} and its extension to {{3D}}.
\newblock \emph{arXiv:2005.05436 [cs, math]}, July 2020.

\bibitem[G(2009)]{allairegAllaire20092d2009}
A.~G.
\newblock Allaire {{G}} (2009) {{A}} 2-d {{Scilab Code}} for shape and topology
  optimization by the level set method.
  {{http://www.cmap.polytechnique.fr/$\sim$allaire/levelset\_en.html}}.
\newblock 2009.

\bibitem[Geuzaine and Remacle(2003)]{gmsh}
C.~Geuzaine and J.-F. Remacle.
\newblock \emph{Gmsh Reference Manual}.
\newblock http://www.geuz.org/gmsh, 1.12 edition, Aug. 2003.

\bibitem[Ghasemi et~al.(2017)Ghasemi, Park, and
  Rabczuk]{ghasemiLevelsetBasedIGA2017}
H.~Ghasemi, H.~S. Park, and T.~Rabczuk.
\newblock A level-set based {{IGA}} formulation for topology optimization of
  flexoelectric materials.
\newblock \emph{Computer Methods in Applied Mechanics and Engineering},
  313:\penalty0 239--258, 2017.

\bibitem[Ghasemi et~al.(2018)Ghasemi, Park, and
  Rabczuk]{ghasemiMultimaterialLevelSetbased2018}
H.~Ghasemi, H.~S. Park, and T.~Rabczuk.
\newblock A multi-material level set-based topology optimization of
  flexoelectric composites.
\newblock \emph{Computer Methods in Applied Mechanics and Engineering},
  332:\penalty0 47--62, 2018.

\bibitem[Hale et~al.(2017)Hale, Li, Richardson, and
  Wells]{haleContainersPortableProductive2017}
J.~S. Hale, L.~Li, C.~N. Richardson, and G.~N. Wells.
\newblock Containers for {{Portable}}, {{Productive}}, and {{Performant
  Scientific Computing}}.
\newblock \emph{Computing in Science Engineering}, 19\penalty0 (6):\penalty0
  40--50, Nov. 2017.
\newblock ISSN 1558-366X.
\newblock \doi{10.1109/MCSE.2017.2421459}.

\bibitem[Hamdia et~al.(2019)Hamdia, Ghasemi, Bazi, AlHichri, Alajlan, and
  Rabczuk]{hamdiaNovelDeepLearning2019a}
K.~M. Hamdia, H.~Ghasemi, Y.~Bazi, H.~AlHichri, N.~Alajlan, and T.~Rabczuk.
\newblock A novel deep learning based method for the computational material
  design of flexoelectric nanostructures with topology optimization.
\newblock \emph{Finite Elements in Analysis and Design}, 165:\penalty0 21--30,
  2019.

\bibitem[Herrero et~al.(2013)Herrero, Mart{\'i}nez, and
  Mart{\'i}]{herreroImplementationLevelSet2013}
D.~Herrero, J.~Mart{\'i}nez, and P.~Mart{\'i}.
\newblock An implementation of level set based topology optimization using
  {{GPU}}.
\newblock In \emph{10th {{World Congress}} on {{Structural}} and
  {{Multidisciplinary Optimization}}, {{Orlando}}, {{Florida}}, {{USA}}}, pages
  1--10, 2013.

\bibitem[Kang and Youn(2016)]{kangIsogeometricTopologyOptimization2016}
P.~Kang and S.-K. Youn.
\newblock Isogeometric topology optimization of shell structures using trimmed
  {{NURBS}} surfaces.
\newblock \emph{Finite Elements in Analysis and Design}, 120:\penalty0 18--40,
  Nov. 2016.
\newblock \doi{10.1016/j.finel.2016.06.003}.

\bibitem[Kharmanda et~al.(2004)Kharmanda, Olhoff, Mohamed, and
  Lemaire]{kharmandaReliabilitybasedTopologyOptimization2004}
G.~Kharmanda, N.~Olhoff, A.~Mohamed, and M.~Lemaire.
\newblock Reliability-based topology optimization.
\newblock \emph{Structural and Multidisciplinary Optimization}, 26\penalty0
  (5):\penalty0 295--307, Mar. 2004.
\newblock \doi{10.1007/s00158-003-0322-7}.

\bibitem[Kirby(2004)]{FIATKirby2004a}
R.~C. Kirby.
\newblock Algorithm 839: Fiat, a new paradigm for computing finite element
  basis functions.
\newblock \emph{ACM Transactions on Mathematical Software}, 30\penalty0
  (4):\penalty0 502--516, 2004.
\newblock \doi{10.1145/1039813.1039820}.

\bibitem[Kirby and Logg(2006)]{FFCKirbyLogg2006a}
R.~C. Kirby and A.~Logg.
\newblock A compiler for variational forms.
\newblock \emph{ACM Transactions on Mathematical Software}, 32\penalty0 (3),
  2006.
\newblock \doi{10.1145/1163641.1163644}.

\bibitem[Lazarov and Sigmund(2011)]{lazarovFiltersTopologyOptimization2011}
B.~S. Lazarov and O.~Sigmund.
\newblock Filters in topology optimization based on {{Helmholtz}}-type
  differential equations.
\newblock \emph{International Journal for Numerical Methods in Engineering},
  86\penalty0 (6):\penalty0 765--781, May 2011.
\newblock ISSN 00295981.
\newblock \doi{10.1002/nme.3072}.

\bibitem[Li et~al.(2010)Li, Kecman, and Salman]{liChunkingMethodEuclidean2010}
Q.~Li, V.~Kecman, and R.~Salman.
\newblock A {{Chunking Method}} for {{Euclidean Distance Matrix Calculation}}
  on {{Large Dataset Using Multi}}-{{GPU}}.
\newblock In \emph{2010 {{Ninth International Conference}} on {{Machine
  Learning}} and {{Applications}}}, pages 208--213, {Washington, DC, USA}, Dec.
  2010. {IEEE}.
\newblock ISBN 978-1-4244-9211-4.
\newblock \doi{10.1109/ICMLA.2010.38}.

\bibitem[Liu and Tovar(2014)]{liuEfficient3DTopology2014}
K.~Liu and A.~Tovar.
\newblock An efficient {{3D}} topology optimization code written in {{Matlab}}.
\newblock \emph{Structural and Multidisciplinary Optimization}, 50\penalty0
  (6):\penalty0 1175--1196, Dec. 2014.
\newblock \doi{10.1007/s00158-014-1107-x}.

\bibitem[Logg and Wells(2010)]{DOLFINLoggWells2010a}
A.~Logg and G.~N. Wells.
\newblock Dolfin: Automated finite element computing.
\newblock \emph{ACM Transactions on Mathematical Software}, 37\penalty0 (2),
  2010.
\newblock \doi{10.1145/1731022.1731030}.

\bibitem[Logg et~al.(2012)Logg, Mardal, Wells, et~al.]{LoggMardalEtAl2012a}
A.~Logg, K.-A. Mardal, G.~N. Wells, et~al.
\newblock \emph{Automated Solution of Differential Equations by the Finite
  Element Method}.
\newblock Springer, 2012.
\newblock ISBN 978-3-642-23098-1.
\newblock \doi{10.1007/978-3-642-23099-8}.

\bibitem[Lu et~al.(2013)Lu, Yamamoto, Otomori, Yamada, Izui, and
  Nishiwaki]{luTopologyOptimizationAcoustic2013}
L.~Lu, T.~Yamamoto, M.~Otomori, T.~Yamada, K.~Izui, and S.~Nishiwaki.
\newblock Topology optimization of an acoustic metamaterial with negative bulk
  modulus using local resonance.
\newblock \emph{Finite Elements in Analysis and Design}, 72:\penalty0 1--12,
  Sept. 2013.
\newblock \doi{10.1016/j.finel.2013.04.005}.

\bibitem[{Mart{\'i}nez-Frutos} and
  {Herrero-P{\'e}rez}(2016)]{martinez-frutosLargescaleRobustTopology2016}
J.~{Mart{\'i}nez-Frutos} and D.~{Herrero-P{\'e}rez}.
\newblock Large-scale robust topology optimization using multi-{{GPU}} systems.
\newblock \emph{Computer Methods in Applied Mechanics and Engineering},
  311:\penalty0 393--414, 2016.

\bibitem[{Mart{\'i}nez-Frutos} and
  {Herrero-P{\'e}rez}(2017)]{martinez-frutosGPUAccelerationEvolutionary2017}
J.~{Mart{\'i}nez-Frutos} and D.~{Herrero-P{\'e}rez}.
\newblock {{GPU}} acceleration for evolutionary topology optimization of
  continuum structures using isosurfaces.
\newblock \emph{Computers \& Structures}, 182:\penalty0 119--136, 2017.

\bibitem[{Mart{\'i}nez-Frutos} and
  {Herrero-P{\'e}rez}(2018)]{martinez-frutosEvolutionaryTopologyOptimization2018}
J.~{Mart{\'i}nez-Frutos} and D.~{Herrero-P{\'e}rez}.
\newblock Evolutionary topology optimization of continuum structures under
  uncertainty using sensitivity analysis and smooth boundary representation.
\newblock \emph{Computers \& Structures}, 205:\penalty0 15--27, 2018.

\bibitem[{Mart{\'i}nez-Frutos} et~al.(2017){Mart{\'i}nez-Frutos},
  {Mart{\'i}nez-Castej{\'o}n}, and
  {Herrero-P{\'e}rez}]{martinez-frutosEfficientTopologyOptimization2017}
J.~{Mart{\'i}nez-Frutos}, P.~J. {Mart{\'i}nez-Castej{\'o}n}, and
  D.~{Herrero-P{\'e}rez}.
\newblock Efficient topology optimization using {{GPU}} computing with
  multilevel granularity.
\newblock \emph{Advances in Engineering Software}, 106:\penalty0 47--62, Apr.
  2017.
\newblock ISSN 0965-9978.
\newblock \doi{10.1016/j.advengsoft.2017.01.009}.

\bibitem[Mlejnek(1992)]{mlejnekAspectsGenesisStructures1992}
H.~P. Mlejnek.
\newblock Some aspects of the genesis of structures.
\newblock \emph{Structural optimization}, 5\penalty0 (1):\penalty0 64--69, Mar.
  1992.
\newblock \doi{10.1007/BF01744697}.

\bibitem[Nanthakumar et~al.(2016)Nanthakumar, Lahmer, Zhuang, Park, and
  Rabczuk]{nanthakumarTopologyOptimizationPiezoelectric2016}
S.~S. Nanthakumar, T.~Lahmer, X.~Zhuang, H.~S. Park, and T.~Rabczuk.
\newblock Topology optimization of piezoelectric nanostructures.
\newblock \emph{Journal of the Mechanics and Physics of Solids}, 94:\penalty0
  316--335, 2016.

\bibitem[Nanthakumar et~al.(2017)Nanthakumar, Zhuang, Park, and
  Rabczuk]{nanthakumarTopologyOptimizationFlexoelectric2017}
S.~S. Nanthakumar, X.~Zhuang, H.~S. Park, and T.~Rabczuk.
\newblock Topology optimization of flexoelectric structures.
\newblock \emph{Journal of the Mechanics and Physics of Solids}, 105:\penalty0
  217--234, 2017.

\bibitem[Nguyen et~al.(2020)Nguyen, Zhuang, Chamoin, Zhao, {Nguyen-Xuan}, and
  Rabczuk]{nguyenThreedimensionalTopologyOptimization2020}
C.~Nguyen, X.~Zhuang, L.~Chamoin, X.~Zhao, H.~{Nguyen-Xuan}, and T.~Rabczuk.
\newblock Three-dimensional topology optimization of auxetic metamaterial using
  isogeometric analysis and model order reduction.
\newblock \emph{Computer Methods in Applied Mechanics and Engineering},
  371:\penalty0 113306, 2020.

\bibitem[\O{}lgaard and Wells(2010)]{FFCOlgaardWells2010b}
K.~B. \O{}lgaard and G.~N. Wells.
\newblock Optimisations for quadrature representations of finite element
  tensors through automated code generation.
\newblock \emph{ACM Transactions on Mathematical Software}, 37, 2010.
\newblock \doi{10.1145/1644001.1644009}.

\bibitem[Ortigosa et~al.(2019)Ortigosa, {Mart{\'i}nez-Frutos}, Gil, and
  {Herrero-P{\'e}rez}]{ortigosaNewStabilisationApproach2019}
R.~Ortigosa, J.~{Mart{\'i}nez-Frutos}, A.~J. Gil, and D.~{Herrero-P{\'e}rez}.
\newblock A new stabilisation approach for level-set based topology
  optimisation of hyperelastic materials.
\newblock \emph{Structural and Multidisciplinary Optimization}, 60\penalty0
  (6):\penalty0 2343--2371, 2019.

\bibitem[Otomori et~al.(2014)Otomori, Yamada, Izui, and
  Nishiwaki]{otomoriMATLABCodeLevel2014}
M.~Otomori, T.~Yamada, K.~Izui, and S.~Nishiwaki.
\newblock {{MATLAB}} code for a level set-based topology optimization method
  using a reaction diffusion equation.
\newblock \emph{Struct Multidiscip Optim}, 51, 2014.
\newblock \doi{10.1007/s00158-014-1190-z}.

\bibitem[P{\'e}rez(2012)]{perezLevelSetMethod2012}
D.~H. P{\'e}rez.
\newblock \emph{Level {{Set Method Applied}} to {{Topology Optimization}}}.
\newblock 2012.

\bibitem[Schlömer et~al.(2019)Schlömer, McBain, Li, Luu, christos,
  Ferrándiz, Barnes, eolianoe, Dalcin, nilswagner, Gupta, Müller, Schwarz,
  Blechta, Coutinho, Beurle, s1291, Shrimali, Cervone, Heister, Langlois, Peak,
  Sharma, Bussonnier, lgiraldi, Jacquenot, Vaillant, Wilson, Gudchenko, and
  Croucher]{nicoschlomer2018}
N.~Schlömer, G.~McBain, T.~Li, K.~Luu, christos, V.~M. Ferrándiz, C.~Barnes,
  eolianoe, L.~Dalcin, nilswagner, A.~Gupta, S.~Müller, L.~Schwarz,
  J.~Blechta, C.~Coutinho, D.~Beurle, s1291, B.~Shrimali, A.~Cervone,
  T.~Heister, T.~Langlois, S.~Peak, S.~Sharma, M.~Bussonnier, lgiraldi,
  G.~Jacquenot, G.~A. Vaillant, C.~Wilson, A.~U. Gudchenko, and A.~Croucher.
\newblock nschloe/meshio 3.3.1, Dec. 2019.

\bibitem[Schmidt and Schulz(2011{\natexlab{a}})]{schmidt2589LineTopology2011}
S.~Schmidt and V.~Schulz.
\newblock A 2589 line topology optimization code written for the graphics card.
\newblock \emph{Computing and Visualization in Science}, 14\penalty0
  (6):\penalty0 249--256, Aug. 2011{\natexlab{a}}.
\newblock \doi{10.1007/s00791-012-0180-1}.

\bibitem[Schmidt and Schulz(2011{\natexlab{b}})]{schmidt2589LineTopology2011b}
S.~Schmidt and V.~Schulz.
\newblock A 2589 line topology optimization code written for the graphics card.
\newblock \emph{Computing and Visualization in Science}, 14\penalty0
  (6):\penalty0 249--256, Aug. 2011{\natexlab{b}}.
\newblock ISSN 1433-0369.
\newblock \doi{10.1007/s00791-012-0180-1}.

\bibitem[Sigmund(1997)]{sigmundDesignCompliantMechanisms1997}
O.~Sigmund.
\newblock On the design of compliant mechanisms using topology optimization.
\newblock \emph{Journal of Structural Mechanics}, 25\penalty0 (4):\penalty0
  493--524, 1997.

\bibitem[Sigmund(2001)]{sigmund99LineTopology2001}
O.~Sigmund.
\newblock A 99 line topology optimization code written in {{Matlab}}.
\newblock \emph{Structural and Multidisciplinary Optimization}, 21\penalty0
  (2):\penalty0 120--127, Apr. 2001.
\newblock \doi{10.1007/s001580050176}.

\bibitem[Sigmund(2007)]{sigmundMorphologybasedBlackWhite2007}
O.~Sigmund.
\newblock Morphology-based black and white filters for topology optimization.
\newblock \emph{Structural and Multidisciplinary Optimization}, 33\penalty0
  (4-5):\penalty0 401--424, 2007.

\bibitem[Sigmund and Maute(2013)]{sigmundTopologyOptimizationApproaches2013b}
O.~Sigmund and K.~Maute.
\newblock Topology optimization approaches.
\newblock \emph{Structural and Multidisciplinary Optimization}, 48\penalty0
  (6):\penalty0 1031--1055, 2013.

\bibitem[Sigmund and
  Petersson(1998)]{sigmundNumericalInstabilitiesTopology1998}
O.~Sigmund and J.~Petersson.
\newblock Numerical instabilities in topology optimization: A survey on
  procedures dealing with checkerboards, mesh-dependencies and local minima.
\newblock \emph{Structural optimization}, 16\penalty0 (1):\penalty0 68--75,
  1998.

\bibitem[Suresh(2010)]{suresh199lineMatlabCode2010}
K.~Suresh.
\newblock A 199-line {{Matlab}} code for {{Pareto}}-optimal tracing in topology
  optimization.
\newblock \emph{Structural and Multidisciplinary Optimization}, 42\penalty0
  (5):\penalty0 665--679, Nov. 2010.
\newblock \doi{10.1007/s00158-010-0534-6}.

\bibitem[Svanberg(1987)]{svanbergMethodMovingAsymptotes1987}
K.~Svanberg.
\newblock The method of moving asymptotes\textemdash a new method for
  structural optimization.
\newblock \emph{International Journal for Numerical Methods in Engineering},
  24\penalty0 (2):\penalty0 359--373, 1987.
\newblock \doi{10.1002/nme.1620240207}.

\bibitem[Talischi et~al.(2012)Talischi, Paulino, Pereira, and
  Menezes]{talischiPolytopMATLABImplementation2012}
C.~Talischi, G.~H. Paulino, A.~Pereira, and I.~F.~M. Menezes.
\newblock Polytop: A {{MATLAB}} implementation of a general topology
  optimization framework using unstructured polygonal finite element meshes.
\newblock \emph{Struct Multidiscip Optim}, 45, 2012.
\newblock \doi{10.1007/s00158-011-0696-x}.

\bibitem[Vemaganti and Lawrence(2005)]{vemagantiParallelMethodsOptimality2005b}
K.~Vemaganti and W.~E. Lawrence.
\newblock Parallel methods for optimality criteria-based topology optimization.
\newblock \emph{Computer Methods in Applied Mechanics and Engineering},
  194\penalty0 (34):\penalty0 3637--3667, Sept. 2005.
\newblock ISSN 0045-7825.
\newblock \doi{10.1016/j.cma.2004.08.008}.

\bibitem[Wadbro and Berggren(2009)]{wadbroMegapixelTopologyOptimization2009}
E.~Wadbro and M.~Berggren.
\newblock Megapixel {{Topology Optimization}} on a {{Graphics Processing
  Unit}}.
\newblock \emph{SIAM Review}, 51\penalty0 (4):\penalty0 707--721, Nov. 2009.
\newblock ISSN 0036-1445.
\newblock \doi{10.1137/070699822}.

\bibitem[Wang et~al.(2003)Wang, Wang, and Guo]{wangLevelSetMethod2003}
M.~Y. Wang, X.~Wang, and D.~Guo.
\newblock A level set method for structural topology optimization.
\newblock \emph{Comput Methods Appl Mech Eng}, 192, 2003.
\newblock \doi{10.1016/S0045-7825(02)00559-5}.

\bibitem[Wei et~al.(2018)Wei, Li, Li, and Wang]{wei88lineMATLABCode2018}
P.~Wei, Z.~Li, X.~Li, and M.~Y. Wang.
\newblock An 88-line {{MATLAB}} code for the parameterized level set method
  based topology optimization using radial basis functions.
\newblock \emph{Structural and Multidisciplinary Optimization}, 58\penalty0
  (2):\penalty0 831--849, Aug. 2018.
\newblock \doi{10.1007/s00158-018-1904-8}.

\bibitem[Zhang et~al.(2016)Zhang, Yuan, Zhang, and
  Guo]{zhangNewTopologyOptimization2016}
W.~Zhang, J.~Yuan, J.~Zhang, and X.~Guo.
\newblock A new topology optimization approach based on moving {{Morphable}}
  components ({{MMC}}) and the ersatz material model.
\newblock \emph{Struct Multidiscip Optim}, 53, 2016.
\newblock \doi{10.1007/s00158-015-1372-3}.

\bibitem[Zhou and Rozvany(1991)]{zhouCOCAlgorithmPart1991}
M.~Zhou and G.~Rozvany.
\newblock The {{COC}} algorithm, {{Part II}}: {{Topological}}, geometrical and
  generalized shape optimization.
\newblock \emph{Computer Methods in Applied Mechanics and Engineering},
  89\penalty0 (1-3):\penalty0 309--336, Aug. 1991.
\newblock \doi{10.1016/0045-7825(91)90046-9}.

\bibitem[Zhou et~al.(2012)Zhou, Cadman, Chen, Li, Xie, and
  Huang]{zhouDesignFabricationBiphasic2012}
S.~Zhou, J.~Cadman, Y.~Chen, W.~Li, Y.~M. Xie, and X.~Huang.
\newblock Design and fabrication of biphasic cellular materials with transport
  properties \textendash{} a modified bidirectional evolutionary structural
  optimization procedure and {{MATLAB}} program.
\newblock \emph{Int J Heat Mass Transf}, 55, 2012.
\newblock \doi{10.1016/j.ijheatmasstransfer.2012.08.028}.

\bibitem[Zuo and Xie(2015)]{zuoSimpleCompactPython2015}
Z.~H. Zuo and Y.~M. Xie.
\newblock A simple and compact {{Python}} code for complex {{3D}} topology
  optimization.
\newblock \emph{Advances in Engineering Software}, July 2015.
\newblock \doi{10.1016/j.advengsoft.2015.02.006}.

\end{thebibliography}

\end{document}